\documentclass[10pt,manuscript,letter]{emulateapj}
\usepackage{appendix}

\newcommand{\chandra}{{\it Chandra}}
\newcommand{\sigT}{\mbox{$\sigma_{\mbox{\tiny T}}$}}
\newcommand{\Tcmb}{\mbox{$T_{\mbox{\tiny CMB}}$}}
\newcommand{\kB}{\mbox{$k_{\mbox{\tiny B}}$}}

\begin{document}

\title{MUSTANG high angular resolution Sunyaev-Zel'dovich Effect imaging of sub-structure
  in four galaxy clusters}

\author{P.M. Korngut\altaffilmark{1,2}, S.R. Dicker\altaffilmark{1},
  E. D.Reese\altaffilmark{1}, B.~S. Mason\altaffilmark{3},
  M.J. Devlin\altaffilmark{1} 
  ,T. Mroczkowski\altaffilmark{1,5}, C. L. Sarazin\altaffilmark{4},
  M.~Sun\altaffilmark{4} and J. Sievers\altaffilmark{6}}

\altaffiltext{1}{University of Pennsylvania, 209 S. 33rd St., Philadelphia, PA 19104, USA}
\altaffiltext{2}{contact author: pkorngut@physics.upenn.edu}
\altaffiltext{3}{National Radio Astronomy Observatory, 520 Edgemont Rd. Charlottesville VA 22903, USA}
\altaffiltext{4}{Department of Astronomy, University of Virginia, P.O. Box 400325, Charlottesville, VA 22904-4325}
\altaffiltext{5}{NASA Einstein Fellow}
\altaffiltext{6}{The Canadian Institute of Theoretical Astrophysics,
  60 St. George Street, Toronto, Ontario M5S 3H8}


\begin{abstract}

We present $10''$ to $18''$ images of four massive clusters of galaxies
through the Sunyaev-Zel'dovich Effect (SZE). These measurements, made
at 90~GHz with the MUSTANG receiver on
the Green Bank Telescope (GBT), reveal pressure sub-structure to the
intra-cluster medium (ICM) in three of the four systems.  We identify the likely presence of a
previously unknown weak shock-front in MACS0744+3927.  By fitting the
Rankine-Hugoniot density jump conditions in a complementary SZE/X-ray
analysis, we infer a Mach 
number of $\mathcal{M} = 1.2^{+0.2}_{-0.2}$ and a shock-velocity of
$1827^{+267}_{-195}$~km~s$^{-1}$.  In RXJ1347-1145,
we present a new reduction of previously reported data and
confirm the presence of a south-east SZE enhancement
with a significance of $13.9\sigma$ when smoothed to $18''$ resolution.  This too is likely caused by
shock-heated gas produced in a recent merger.  In our highest redshift system,
CL1226+3332, we detect sub-structure at a peak significance of $4.6\sigma$
in the form of a ridge oriented orthogonally to the vector
connecting the main mass peak and a sub-clump revealed by weak
lensing.  We also conclude that the gas distribution is elongated in a
south-west direction, consistent with a previously proposed merger
scenario.  The SZE image of the cool core
cluster Abell 1835 is, in contrast, consistent with azimuthally
symmetric signal only.  This pilot study 
demonstrates the potential of high-resolution SZE images to complement X-ray data
and probe the dynamics of galaxy clusters.  
\end{abstract}


\section{Introduction}
\label{sec:intro}

The Sunyaev-Zel'dovich Effect (SZE) in clusters of galaxies arises
from inverse Compton scattering of cosmic microwave background (CMB) photons
off hot electrons in the Intra-Cluster Medium (ICM) \citep{sz72}.  The magnitude
of this effect is redshift independent and directly proportional to
the line of sight 
integrated pressure of the plasma. At frequencies $\lesssim 218$ GHz, it is manifested as a
decrement in CMB intensity.
Over the past two decades, measurements of the SZE in
clusters of galaxies have been used to probe a wide range of cosmological and
astrophysical questions. It has been used by dedicated surveys to
search for clusters \citep[e.g.,][]{actclus,sptcat,actclus2,toby10}, combined with X-ray
data to measure the Hubble flow
\citep[e.g.,][]{brianho,reese02,bonamente06} and to derive
physical cluster properties from radial profiles \citep[e.g.,][]{laroque03,bonamente06,tony09}. For
reviews of the SZE and its applications, see \citet{birk99} and \citet{chr}.

Measurements of the SZE at high angular resolution are difficult
because of the large apertures required.     
Nearly all measurements currently in the literature have effective
angular resolution larger than $\sim 1'$.  These angular scales
(corresponding to $\sim 365$ kpc at $z = 0.5$) are extremely useful for measuring
the bulk signal out to large cluster-centric radii \citep[e.g.,][]{nord09} but are unable to
resolve the smaller scale physical processes in the cluster cores.

High-resolution X-ray imaging from \chandra\ and XMM-Newton in the
last decade opened a new window to cluster physics.  Objects once thought to 
be spherically symmetric and relaxed have been shown to
display evidence of interesting phenomena which provide insight to the
complicated dynamics at play in these structures. Among these are
shocks and cold-fronts induced by recent mergers \citep[e.g.,][]{markshocks}, cavities and
heating caused by AGN interactions \citep[e.g.,][]{mcbubbles} and sharp surface
brightness edges caused by gas sloshing \citep[e.g.,][]{sloshing}. 

High-resolution images of the SZE in clusters provide a new tool
which, when combined with X-ray measurements, can constrain complicated physics
in galaxy clusters. This is particularly true of the high-redshift
Universe as the X-ray surface brightness data (proportional to the
product of the density squared and square root of temperature
integrated along the line of sight) suffer from cosmological dimming. 

The potential of resolved SZE was first 
demonstrated by \citet{komatsu01}, who used the Nobeyama 45m to image
RXJ1347-1145, a massive X-ray luminous cluster previously thought to be
relaxed and spherically symmetric \citep{schindler97}.  The asymmetry
revealed by their work was the first indication that the system was
disturbed, and it is now believed to have undergone a recent merger
\citep{kitayama04}. This has since been confirmed in the SZE by MUSTANG
\citep{mason10}.
More recently, other groups have made high
resolution SZE images in  CL J0152-1347
\citep[][at $\sim35''$ resolution]{massardi10} and the Bullet Cluster
\citep[][at $\sim 30''$ resolution]{malu10}. 

In this work, we present measurements taken with
the Multiplexed SQUID/TES Array at Ninety Gigahertz (MUSTANG)
receiver on the  100m Robert C. Byrd Green Bank Telescope (GBT).
The large collecting area of the GBT
combined with the focal plane array of bolometers
make this system ideal for probing sub-structure in clusters through the
SZE.  

The clusters MACS0744, RXJ1347, and CL1226
were selected because they displayed signs of merger activity in
previous measurements and therefore were likely to contain small-scale
features created by 
merger processes. In contrast, Abell 1835 was specifically targeted
because it was expected to be representative of a typical relaxed cluster.
All uncertainties quoted in this paper are $68\%$ confidence and we
assume a cosmology where $H_{0}=71$ km s$^{-1}$ Mpc$^{-1}$,
$\Omega_{\Lambda}=0.73$ and $\Omega_{M}=0.27$.

\section{Instrument \& Observations}
\label{sec:obs}
\subsection{MUSTANG}
MUSTANG is
a focal plane camera with an
$8\times8$ array of Transition Edge Sensor (TES) bolometers built for
the Gregorian focus of the 100 m Robert C. Byrd Green Bank Telescope (GBT).
It has 18.4~GHz of bandwidth centered
on 90 GHz.  The array has a $0.63 f\lambda$ pixel spacing which yields a well sampled
instantaneous field of view (FOV) of $42''$ on the sky. 
More detailed information on the instrument can be found on the MUSTANG website\footnote{{\tt
http://www.gb.nrao.edu/mustang/}} and in \citet{spie08,orion}.

\subsection{Observations}
\begin{figure}
\plotone{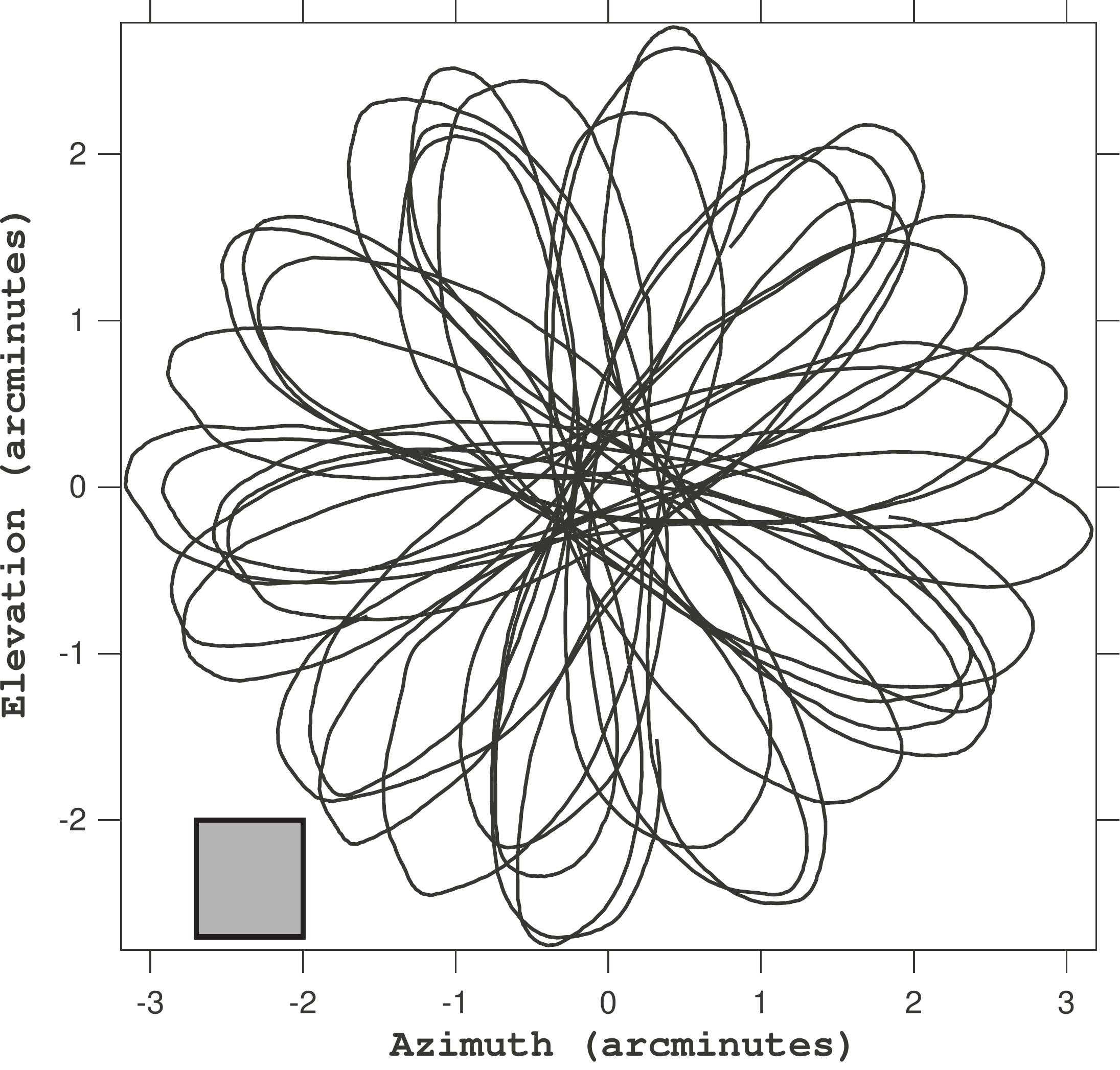}
\caption{GBT bore-sight trajectory during a Lissajous daisy scan
  pattern. This observing strategy was used for the majority of the
  observations presented.  The movement of the source across the sky
  during the scan has been subtracted.  The shaded box indicates the
  instantaneous field of view of MUSTANG}
\label{figure:dais}
\end{figure}

Data presented here were obtained during the winter/spring of 2009 and 2010.
The cluster signal was modulated 
predominantly in a ``Lissajous daisy'' scan pattern.  This strategy was
designed to move the telescope with high speed ($\sim
0.5'$s$^{-1}$) without drastic accelerations which can induce 
feed arm instabilities and pointing wobble.  Faster scan rates move
the sky signal to frequencies above the low frequency ($1/f$) noise
from the atmosphere and internal fluctuations.  The GBT bore-sight
trajectory during one of these scans is displayed in Figure
\ref{figure:dais}.  This observation pattern is centrally weighted and produces
maps with radially increasing noise levels. While the particular scan
shown in Figure \ref{figure:dais} contains information in a
$\sim6'$ diameter region, only the central $\sim2'$ are well covered.
 To improve the sky sampling in cluster
cores, each object was mapped with 5 tiled pointing centers: one centered on
the X-ray surface brightness peak and others offset to the north,
south, east and west by approximately one instantaneous field of view (FOV) ($\sim
42''$).  Other scan patterns with more uniform sky coverage 
such as the ``billiard ball'' scan described in
\citet{orion}, \citet{agn} and \citet{mason10} were used as well.  This alternative strategy has the
advantage of uniform noise across the map, but at the cost of slower telescope
velocity with sharper turnarounds.

At the start of each session, out-of-focus (OOF) holography was
carried out using a bright ($\sim1$ Jy) unresolved source.  This
technique, described in detail in \citet{bojanoof}, consists of mapping
a compact source with the GBT secondary in three positions relative to
the primary: nominally in focus and $3\lambda$ on either side. An
automated real-time analysis uses the measured beam patterns
to fit for phase errors in the
telescope aperture.  Corrections to the active surface of the GBT primary
are calculated and applied along with pointing and focus offsets.

\begin{deluxetable}{lccc}
\tablecaption{Observation Summary \label{tbl:obs}}
\tablehead{
Cluster           & $z_r$   & Time    & Secondary Calibrator      \\ 
                  &     & (h)     &           }
\startdata 
Abell 1835        &0.25 & 3.5     & 1415+1320 \\
Rx~J1347.5-1145   &0.45 & 3.3     & 1337-1257 \\
MACS~J0744.8+3927 &0.69 & 5.8     & 0824+3916 \\
CL~J1226.9+3332   &0.89 & 4.5     & 1159+2914 \\
\enddata 
\end{deluxetable}

After every two Lissajous daisy scans ($\sim30$ minutes on source)
the beam profile was measured using a nearby bright compact quasar.  If significant
ellipticity or gain decrease is detected in
the periodic beam measurements, the OOF procedure is repeated. These beam maps are used in image
reconstruction to track fluctuations in the telescope gain, atmosphere and
pointing offsets. The $\sim30$ minute calibration timescale was chosen
as it is characteristic of the thermal time constant of the
telescope. The sources
used as secondary calibrators for each cluster along with the total
on-source integration times are
presented in Table \ref{tbl:obs}.  

Planets were used for absolute flux calibration and were mapped at least
once per night. The fluxes of these primary calibrators are taken from
\citet{weiland10}.  Several times a night, off-source scans with the telescope
at rest and the internal calibration lamp (CAL) firing with a 0.5 Hz square
wave were taken.  These are used in analysis to fit for the gain of each pixel.
The absolute flux of the data is calibrated to an accuracy of $15\%$.

\section{MUSTANG Data Reduction}
\label{sec:data}
A custom imaging algorithm implemented in IDL is used to produce maps
from the time ordered bolometer data.  The data is heavily filtered to
remove atmospheric signal prior to map making.
The process is outlined below: 

\begin{enumerate}
\item Gain inhomogeneities across the detector array are flat-fielded
  using the nearest CAL scan. These data are also used to identify and
  mask unresponsive pixels (typically $10$-$15$ out of
  64). 

\item A template of the atmospheric signal is estimated from
  low frequency fluctuations that are highly correlated across the
  array.  This is constructed from an average of the time streams from all
  accepted pixels.  The model is then low-pass filtered in Fourier
  space to separate the astronomical signal on small spatial scales from the
  atmospheric template. This filtering requires a characteristic
  frequency based on the noise properties of the data. The template is then
  subtracted  in the time domain from each
  pixel.  The effectiveness of this filter relies on the assumption
  that the celestial
  signal is not common mode, which is valid only in the limit of
  compact sources. Bulk signals from clusters are not well approximated
  in this assumption.  It is therefore essential
  to simulate and quantify the angular transfer function of the
  imaging pipeline. 

\item A low-order polynomial is fit and subtracted from each
  time stream.  This further removes the long timescale fluctuations
  in the data.

\item The data all contain a coherent 1.411 Hz signal. This is produced by
  fluctuations in optical load on the detectors caused by the thermal
  cycle of the camera's main cryogenic refrigerator.   It is well
  approximated by a sinusoid and is 
  removed at this stage.

\item A per-pixel high pass filter is applied in Fourier space.  This
  aggressive technique removes all low frequency spatial modes from the data
  indiscriminately.  A characteristic
  frequency is defined at this stage as well. This will further affect
  the angular scales present in the reconstructed image.

\item Individual detector weights are computed based on the noise
  characteristics of each detector after the processing described
  above. This is used to create an effective exposure time for each
  pixel on the sky.

\item The time stream data are then binned on a
  $2''\times2''$ grid in Right Ascension and Declination. 

\end{enumerate}


\begin{figure}
\plotone{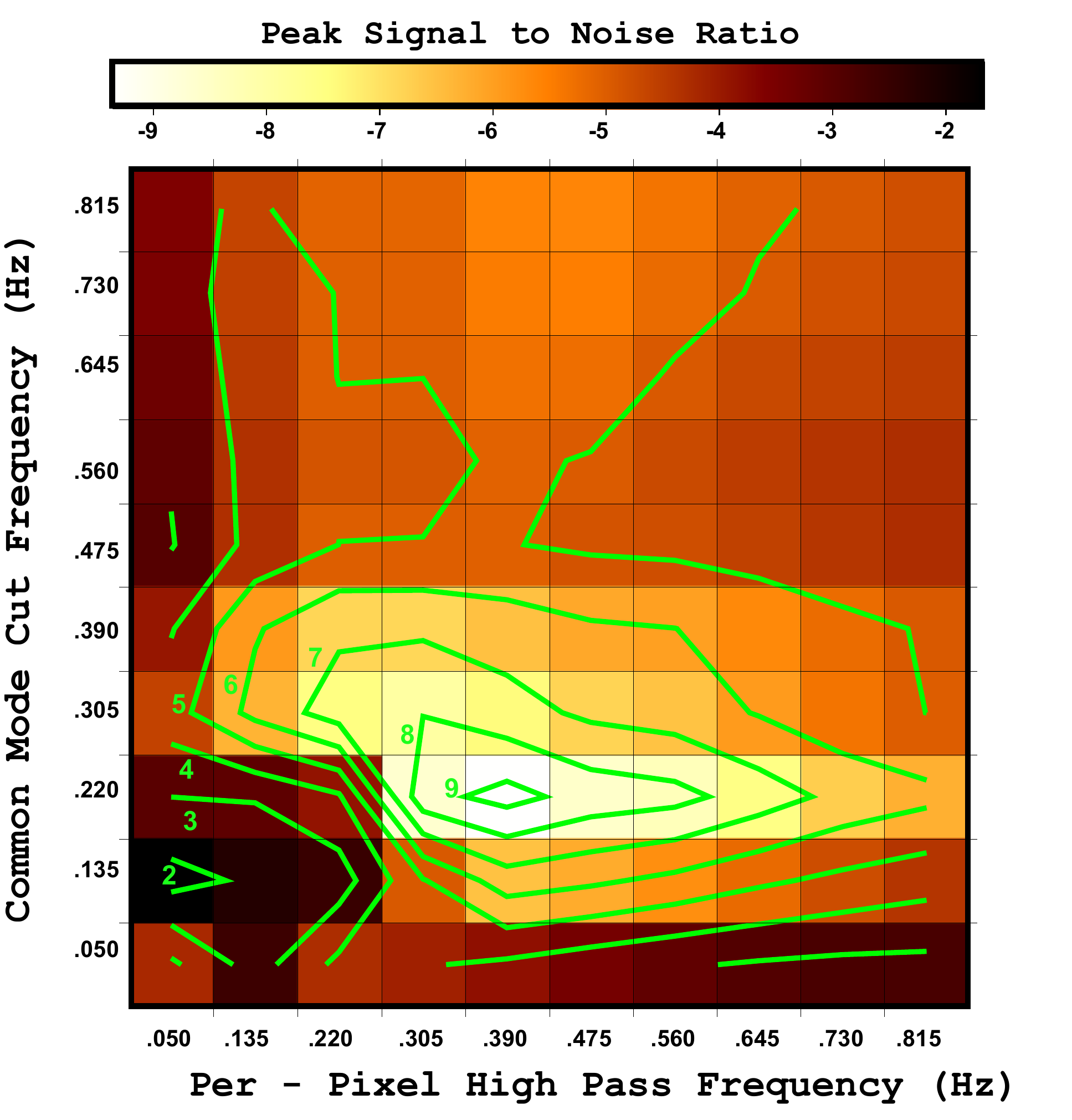}
\caption{Parameter space optimization for the CL1226+3332 data.
  Color scale conveys the peak signal to noise in the exposure corrected map
  convolved with a $10''$ Gaussian produced with each filter parameter combination.}
\label{fig:opt}
\end{figure}


The MUSTANG images presented in this work have been optimized for peak signal
to noise on the compact features of the clusters.  As described above,
there are two selectable filter parameters used in the map maker, one
for the
common mode template and the other for the per-pixel high-pass.  These selected frequencies
correspond to spatial scales on the sky through the speed at which the
signal is modulated by the telescope scan (usually $\sim 0.5'$s$^{-1}$).  The optimal filter for each
object depends on the intrinsic structure of the source as well as the
noise properties of the scans 
used in each observation.  To determine the
optimal filter for each map, parameter space is explored systematically by
mapping each object with varied degrees of filtering. The peak
signal-to-noise ratio (SNR) as a function of these two parameters is displayed in Figure
\ref{fig:opt}. SNR decreases towards the top right of this figure
because too much signal is being filtered out. It also decreases towards the lower left
because too much atmosphere is allowed in the map. A single optimal value for each parameter is assumed
for the entire data set on each cluster.  

The noise in each map is defined by the standard
deviation of all pixels in an off-source region free from obvious
signal.  From extensive Monte-Carlo simulation, we find that this
calculation provides a good measurement of the noise in the map as a
whole, provided that it is scaled by the square root of the difference
in the map weights of the areas in question.

The necessary filtering steps described above result in an attenuation
of flux in the recovered map.  The magnitude of this attenuation
depends strongly on angular scale.  Typically, all structure larger
than $\sim 1.5\times$ the instantaneous FOV is removed entirely.  The
angular transfer function for each
object mapped is calculated using the specific scans and filter
parameters selected to produce the map.  When quantitatively comparing
model images to observed data, it is essential to
apply this transfer function to the model before doing so.  A more
detailed description of the calculation of the transfer function is
described in \citet{mason10}.

\section{\chandra\ data reduction}
\label{sec:chandata}
Archival \chandra\ data are reduced using CIAO version 4.2 and
calibration database 4.2.0.  Starting with the level 1 events file,
standard corrections are applied along with light curve filtering and other
standard processing \citep[for reduction details see,
][]{reese2010}.  Images are made in full resolution ($
0\farcs492$ pixels) and exposure maps are computed at 1 keV.  When
merging data from separate observations, 
images and exposure maps from each data set are combined and a
wavelet based source detector is used on the combined image and
exposure map to find and generate a list of potential point sources.
The list is examined and adjusted by eye and used for our point source
mask.  A summary of the archival \chandra\ data used in this paper is
presented in Table \ref{tbl:xobs}

\begin{deluxetable}{lcc}
\tablecaption{Archival {\it Chandra} Data \label{tbl:xobs}}
\tablehead{
Cluster           &  Time  & ObsIds           \\ 
                  & (ksec) &           }
\startdata 
Abell 1835        & 222    & 495, 496, 6880   \\
                  &        & 6881, 7370       \\

MACS~J0744.8+3927 & 90     & 3197, 3585, 6111 \\
CL~J1226.9+3332   & 74     & 3810, 5014, 932  \\
\enddata 
\end{deluxetable}

\section{Results}
\label{section:results}

\subsection{MACS J0744+3927 (z = 0.69)}
This massive high-redshift system, found in the MAssive Cluster Survey
(MACS) of the all sky ROSAT data \citep{ebeling01_MACSfind}, has appeared in several
studies using X-ray and SZE data
\citep[e.g.,][]{laroque03,laroque06,ebeling07_12MACS}.  Unlike the
other clusters in our sample, targeted multiwavelength studies of
MACS0744 are scarce in the literature. \citet{kartaltepe08} include
this object in a red sequence galaxy distribution study of a sub-sample of 12
MACS clusters. They note that understanding the assembly dynamics of
this system is made difficult by its complex morphology, which
includes some evidence
of a dense core in the X-ray images, and an elongated doubly peaked
distribution of red sequence galaxies. 

Indications of a hot component are
present  in the literature as well, particularly
in \citet{laroque03} who imaged this system in SZE on arcminute scales with the
BIMA/OVRO telescopes.  Assuming a gas mass fraction within
$r_{500}$\footnote{The radius within which the density is 500 times that of the critical
density of the Universe.} of
$f_g=0.081^{+0.009}_{-0.011}$, they obtained a best fit SZE temperature of
$k_BT_{e}=17.9^{+10.8}_{-3.4}$ keV.  The indications of hot gas, with a high
central density, in an object at high-redshift compelled us to include this cluster in our sample.
It also has the favorable characteristic of having no known radio sources
in close proximity on the sky.

\subsubsection{MUSTANG Data}
\label{subsubsec:m0744_mustang}
The SZE map produced from 5.8 hours of MUSTANG data is shown in Figure
\ref{fig:m0744money}. It
consists of a kidney shaped ridge $\sim25''$ long in the north-south
direction. From east to west, the structure is roughly the width of
our beam, and thus is not resolved in this direction.  The curvature of this
feature is well described empirically as an 80 degree sector of an ellipse 
with an axial ratio of $1.25$, with the minor axis and center of the
observed SZE being $12$ degrees south of west on the sky.  

\subsubsection{\chandra\ Data}
\label{subsubsec:m0744_chandra}
The \chandra\ image is shown beside the MUSTANG map in Figure
\ref{fig:m0744money}.  It was produced from nearly 90 ksec of combined archival data
merged from ObsIDs 3197, 3585 and 6111 and
reduced with the method described in section \ref{sec:chandata}. 
The core of this cluster displays an
asymmetric X-ray surface brightness morphology with a sharp
discontinuity on the western edge.
The concave side of the SZE peak identified by MUSTANG is
aligned concentrically with the convex edge of the surface brightness discontinuity in
the X-ray.  Such an enhancement in the SZE in a location offset from the peak in X-ray
surface brightness requires a significantly heated plasma.

\subsubsection{X-ray Surface Brightness Shock Modeling}
\label{subsubsec:m0744_shockmodel}

\begin{figure*}
\plotone{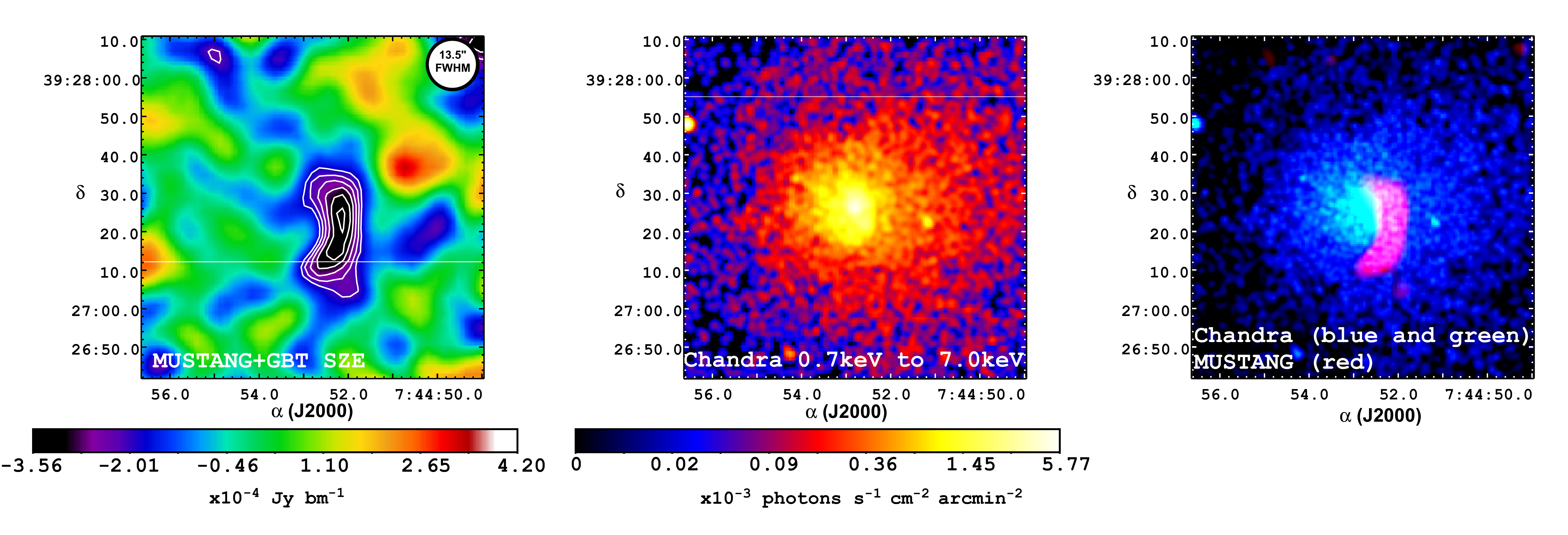}
\caption{SZE and X-ray images of MACS0744. Left: MUSTANG+GBT SZE at $13\farcs5$ FWHM effective resolution
  after  smoothing.  Contours are multiples of $0.5 \sigma$ starting
  at $3\sigma$. Center: Chandra X-ray surface brightness in the
  cluster core.  The image has been smoothed with a $1\farcs5$
  Gaussian. Right: Composite image of Chandra X-ray and MUSTANG
  SZE.  Blue and Green are identical data on different logarithmic
  color scales.  Red shows the MUSTANG SZE data.  The kidney shaped
  ridge revealed by MUSTANG is aligned concentrically with a sharp
  surface brightness discontinuity in the Chandra map.}
\label{fig:m0744money}
\end{figure*}

\begin{figure}
\epsscale{0.7}
\plotone{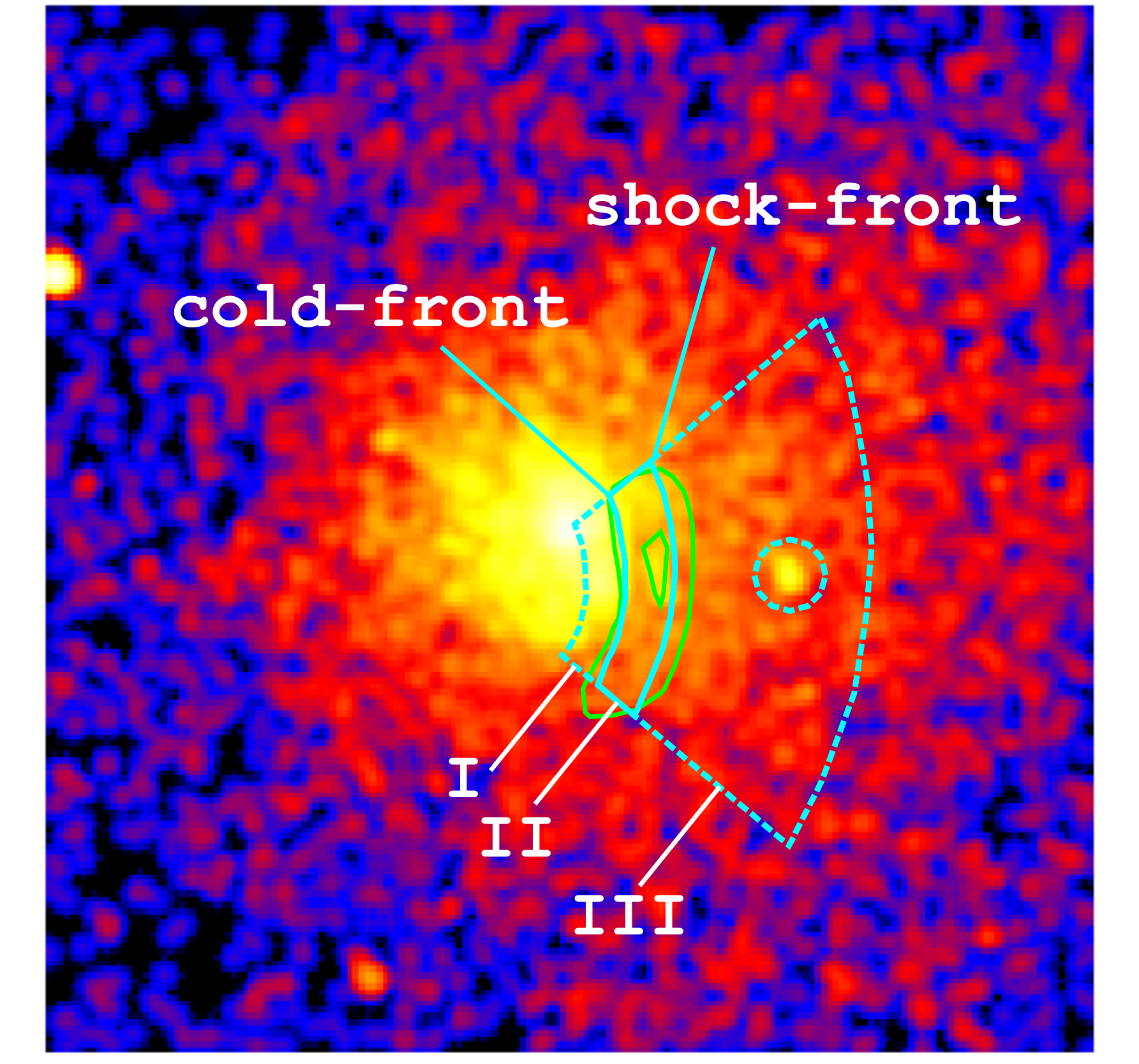}
\caption{Geometry and regions used for elliptical profiles and
  X-ray spectroscopy on MACS0744 overlaid on the \chandra\ surface Brightness
  image. Green contours are $(-4.5,-5.5)\sigma$ SZE decrement. The
  three regions correspond to the cool intact core (I), the shock
  heated gas (II) and pre-shock region (III). One X-ray point
  source has been excised from the pre-shock region. The borders of
  the wedge indicate the azimuthal range used in producing radial profiles.}   
\label{fig:shock0744}
\end{figure}

The combined SZE and X-ray image morphology presented in
Figure \ref{fig:m0744money} is suggestive of a system dominated by a
merger driven shock-front. Arriving at this conclusion based on the
existing relatively low SNR X-ray and SZE data alone would be quite
tenuous; however, 
the kidney shaped ridge seen by MUSTANG combined with the sharp edge seen by
\chandra\ is difficult to explain without invoking a shock-heating mechanism. 
We proceed to model the system in the framework of a shock-front through
a complementary analysis of X-ray and SZE in the approach outlined below:

\begin{itemize}

 \item The elliptical geometry and location of the shocked gas is approximated from the SZE data.

 \item This geometry is used to fit a two dimensional X-ray surface brightness
   profile with a model consisting of three regions: a cold intact
   core bordered by a cold-front, a shock-heated region bordered by a
   shock-front, and a pre-shock region.  These
   correspond to I, II and III, respectively, in Figure \ref{fig:shock0744}.

 \item X-ray spectroscopy is performed in each region to obtain the plasma temperature.

 \item Three dimensional density and pressure models are produced from
   the surface brightness and spectral fits.
 
 \item The pressure model is integrated along the line of sight to
   produce a two dimensional Compton $y_C$ map.

 \item A mock SZE image is constructed at the resolution and with the angular
   extent of the MUSTANG map, and the model and data are compared.
\end{itemize}

We model the X-ray emissivity as a
power law, $\varepsilon \propto r^{-p}$, within each region assuming
an  ellipsoidal geometry with two axes in the plane of the sky
and one along the line of sight (see Appendix~\ref{section:shock} for
details).  The model has 8 parameters in total, two characteristic
radii, and a normalization and power law
index in each of three regions.  We perform a Markov chain Monte Carlo (MCMC) analysis using
Poisson statistics for the X-ray data \citep[for analysis and
  statistics details see, e.g., ][]{reese2000, reese02, bonamente06}.
Each chain is run for a million iterations.  Convergence and mixing
are checked by running two chains and comparing them against one
another \citep{gelman1992, verde2003}.  The choice of burn-in period
does not significantly affect the results but for concreteness we
report results using a burn in of 10,000 iterations.  The model fit is
limited to a wedge subtending 80$^\circ$ and extending from
10\arcsec\ 
to 40\arcsec\ from the nominal center.  This region corresponds
to the region of interest suggested by the SZE and X-ray data as
discussed in Section~\ref{subsubsec:m0744_chandra}.

Initial attempts to model all 8 parameters at once were unsuccessful
due to low SNR in these small regions, with the
chains showing poor convergence.  To limit the number of free
parameters, we implement chains to determine the discontinuity radii,
$R_{s1}$ and $R_{s2}$, individually and then fix those radii.  This
entails using a single discontinuity model, which has 5 parameters,
rather than 8.  The inner discontinuity radius, $R_{s1}$, is
determined with single discontinuity chains using the entire fitting
region.  The outer discontinuity radius, $R_{s2}$, is fit with a
single discontinuity model limiting the fitting region to larger radii
than $R_{s1}$.  

\begin{figure*}[h!]
\plotone{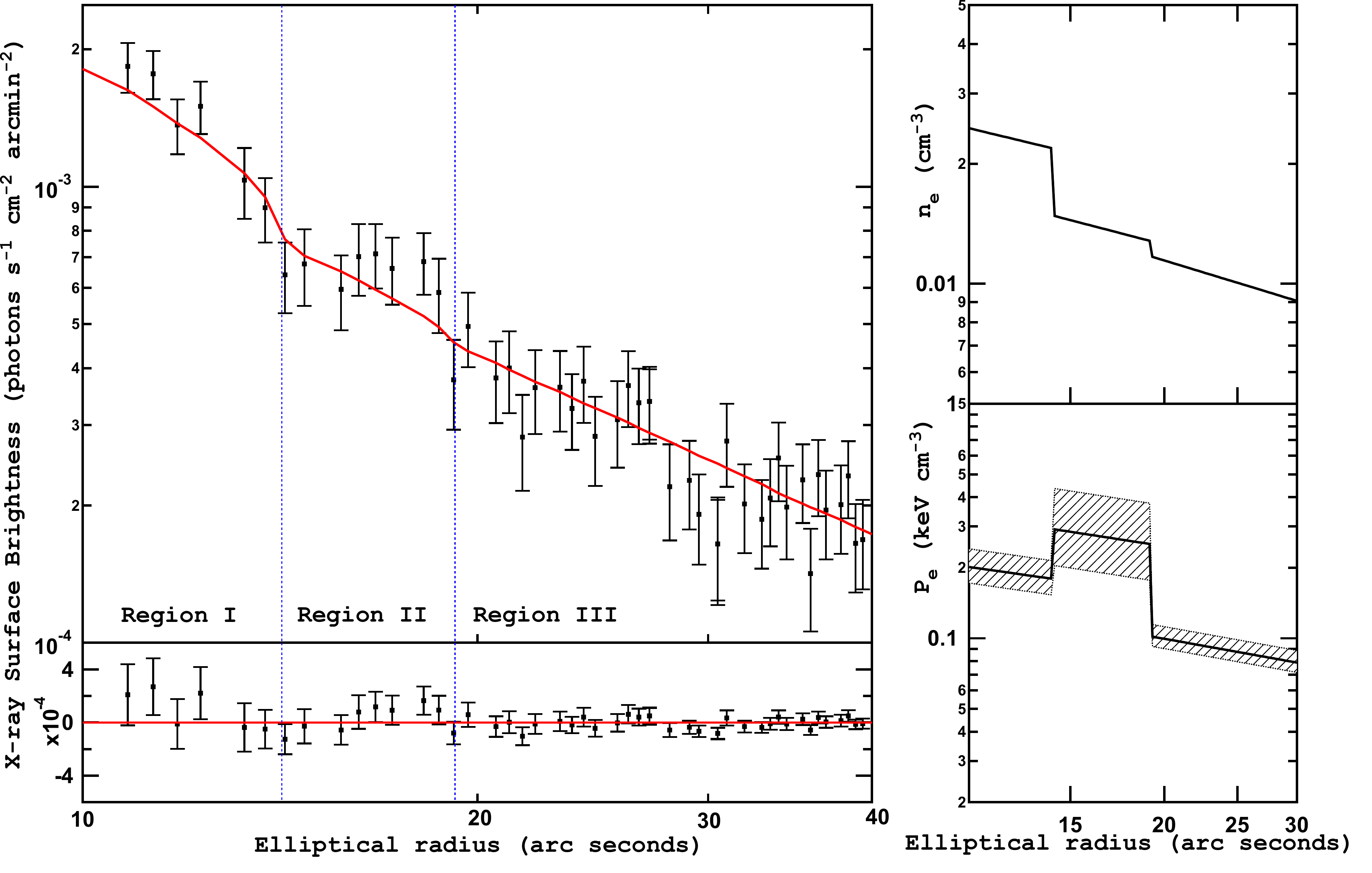}
\caption{Top Left:
  Chandra X-ray surface brightness elliptical profile (points) and best fit
  analytical model in MACS0744 (red line). Bottom left: Residual of
  data and model in top left.  Blue lines show the best fit
  characteristic radii for the cold-front and shock-front, $14\farcs19$
  and $19\farcs23$ respectively.  Top Right: Intrinsic electron number
  density model produced from
  the surface brightness fit on the left.  Bottom Right: Pressure model produced from
  the above density model and the temperatures derived from \chandra\
  spectroscopy.  The shaded regions show the uncertainty based on the
  spectroscopically measured temperatures.  Radii in this figure are elliptical
  and follow the conventions described in Appendix \ref{section:shock}.}
\label{fig:m0744profs}
\end{figure*}

With both discontinuity radii in hand, the double discontinuity model
chains are run with fixed characteristic radii.  This is enough of a
reduction of parameter space to produce converged chains.  Best fit
and 68\% confidence level uncertainties are shown in
Table~\ref{tbl:fitparams}. In this table, the parameter $f$
is defined to be the ratio of the normalization of a given region over
the normalization in the cool intact core (region I).  Because the radial
dependence of the model follows a power law with an 
exponent less than zero, the amplitudes quoted here are normalized at
the cold-front radius, $R_{s1}=14\farcs19$, to avoid a singularity at the
origin. Figure~\ref{fig:m0744profs} shows the X-ray surface
brightness profile within the fitting region along with the best fit
model.

\begin{figure*}[h!]
\plotone{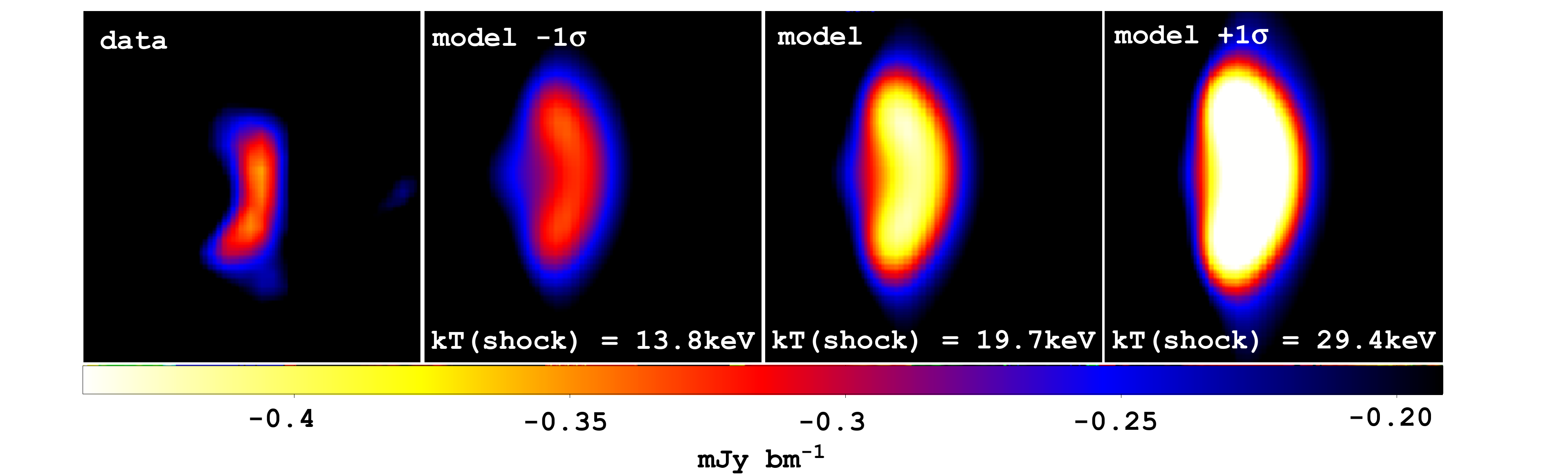}
\caption{MUSTANG SZE data and models in MACS0744. Model images were
  generated by
  integrating the best fit and $1\sigma$ three dimensional pressure
  models from the X-ray along the line
  of sight and passed through the relavent angular transfer function.  Comparison to the MUSTANG data shows the excellent agreement
  with predicted flux scale.  A shock temperature closer to the
  $-1\sigma$ value is favored by the measured SZE data.}
\label{fig:m0744model}
\end{figure*}

\begin{figure*}
\plotone{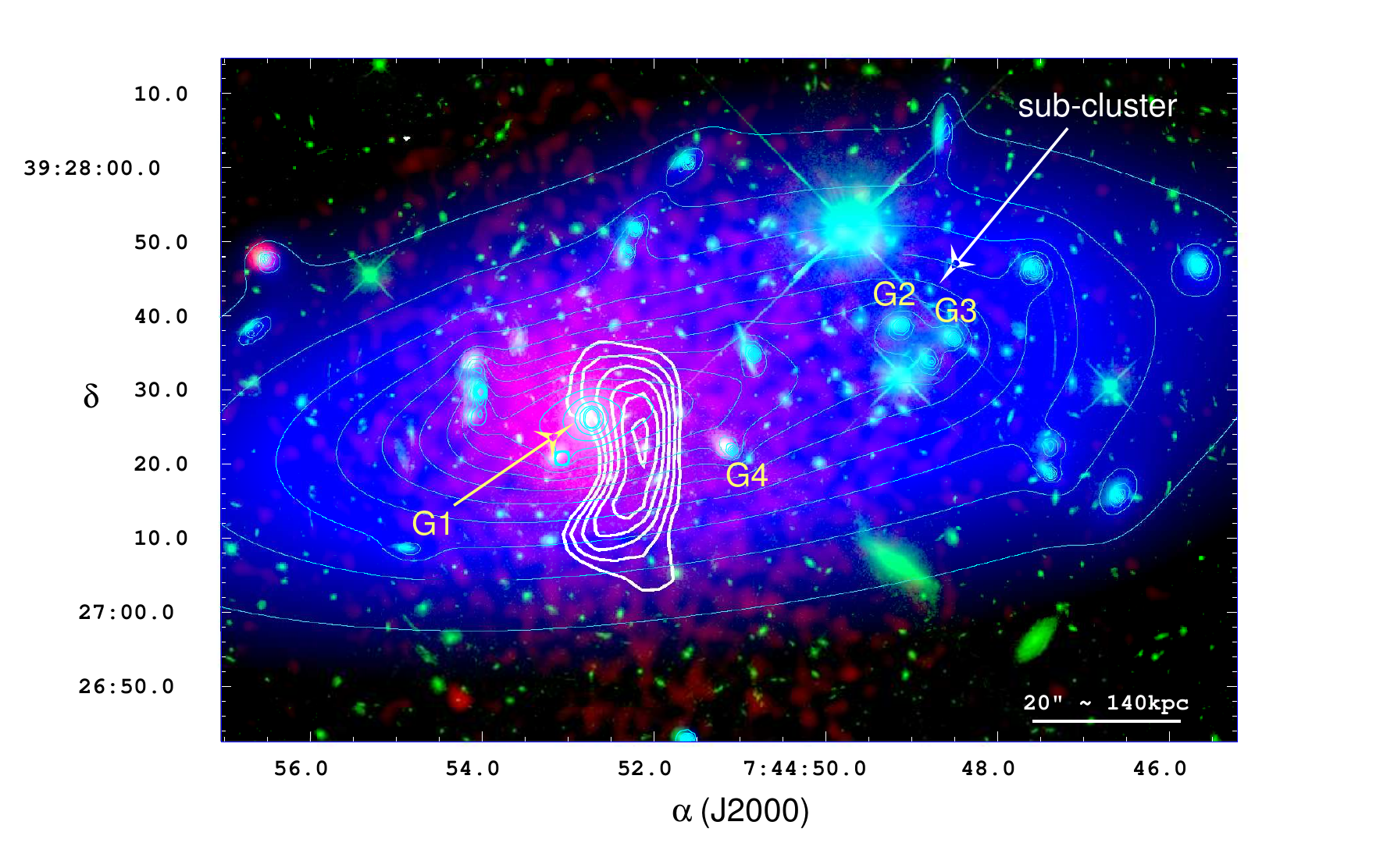}
\caption{Multi-wavelength composite image of MACS0744.  Green is
  HST/ACS data in the F814W band.  Red is \chandra\ X-ray smoothed with
  a $1\farcs5$ Gaussian. Blue color-scale and contours show the strong lensing mass
  reconstruction of \citet{richardlens}. White contours are the
  MUSTANG SZE and are identical to those in Figure
  \ref{fig:m0744money}.  They are in units of SNR to account for uneven exposure across the
  field shown here. Galaxy ``G1'' is the BCG of the main cluster.
  The lensing mass reveals a distinct elongation
  towards the west.  Galaxies ``G2'' and ``G3'' are bright red
  ellipticals located in the
  center of a secondary mass peak with no corresponding baryonic
  emission seen in X-ray.  The SZE shows no enhancement at this
  location either; however, the constraint is weaker as the SZE map
  has large uncertainty at this location due to central weighting of
  scan strategy. ``G4'' is another bright cluster member which harbors an
  X-ray point source.  It too is coincident with a dark matter peak.
  The presence of peaks in mass distribution with no corresponding
  baryons is suggestive of a merger scenario in which an infalling
  sub-cluster has passed through the main core, losing its baryons to
  ram pressure stripping.  It is likely that the weak shock identified
  by MUSTANG was produced by one of these events.}
\label{figure:m0744composite}
\end{figure*}

We also ran MCMC fits modeling a constant X-ray background in addition
to the shock model.  It has no statistically significant effect on the
shock model results.  This is not surprising as the X-ray background
is over an order of magnitude down in surface brightness compared to
the cluster signal at the outermost radius considered in the fit.  The
X-ray background becomes even less important towards the inner radii
where the cluster signal rises.

To produce Compton $y_C$
maps, the three dimensional pressure model was
numerically integrated along the line of sight using Equation
\ref{equation:y} out to an elliptical radius of
$60''$, where the single power law model becomes a poor description of
the X-ray data. This map is then used to produce a predicted SZE image at
90 GHz. After convolving with the GBT beam, the angular transfer
function of the analysis pipeline is applied to the model in Fourier space and
compared to the measured SZE data. Since the model is only valid in a
specified range of angles about the center of the ellipse, the remaining sky was assumed to be well
described by the double $\beta$ model of \citet{laroque06} and a
single temperature of $8.0$ keV. Three model MUSTANG maps were
produced using this process and are shown alongside the data in
Figure \ref{fig:m0744model}.  The model uncertainty is dominated by
the errors in spectroscopic $k_BT_e$ in region II. To account for this in
data comparison we show three model images corresponding to
pressure models produced with the best fit and the temperature fits to
\chandra\ data at
$\pm1\sigma$.   The flux scale in the MUSTANG map is
completely consistent with the X-ray analysis and is
suggestive of a temperature closer to the low end of the allowed
$1\sigma$ parameter space.
\begin{deluxetable}{lccc}
\tablecaption{Best Fit Parameters for the Shock Model in MACS0744 \label{tbl:fitparams}}
\tablehead{
Region & $f$                       & $p$                       &  $k_{B}T_{e}$ \\ 
       &                           &                           & (keV) }
\startdata
I      & 1                         &$ 0.913^{+0.379}_{-0.285}$   & $8.2^{+1.6}_{-1.2}$  \\
II     &$0.480^{+0.124}_{-0.084}$    &$0.986^{+0.559}_{-0.349}$    & $19.7^{+9.7}_{-5.9}$    \\
III    & $0.406^{+0.086}_{-0.063}$   &$1.151^{+0.041}_{-0.040}$    &$8.7^{+1.1}_{-0.8}$   \\ 
\enddata 
\end{deluxetable}

\subsubsection{\chandra\ Spectroscopy}
\label{subsubsec:m0744_chandra_xspec}

Informed by the shock modeling of the \chandra\ data, regions
corresponding to the core, shock heated and pre-shock regions are constructed and used
for spectral extraction.  These correspond to regions I, II and III in
Figure \ref{fig:shock0744}.  Since the calibration varies both in time
and over the ACIS chips, spectra are extracted and response files
computed for each of the 3 observations individually.  All three
spectra are then fit simultaneously.  

XSPEC \citep{arnaud1996, dorman2001} is used to model the ICM with a
Mekal spectrum \citep{mewe1985, mewe1986, liedahl1995, arnaud1985,
  arnaud1992}.  In this fit we account for Galactic extinction and
assume the solar
abundances of \citet{asplund2009}. The cross sections of
\citet{balucinska1992} with an updated He cross section
\citep{yan1998} are used.  The ``cstat'' statistic, which is similar to the \citet{cash1979}
statistic, is used when modeling the data to properly account for low counts.  
All three spectra are fit
simultaneously to the same plasma model with the abundance fixed to be
0.3 solar in all cases.  The normalizations are allowed to float
between data sets.  The fit is limited to photons within the energy
range 0.7-7.0 keV.  Best fit values for the electron temperature and 68\% confidence
ranges are summarized in Table~\ref{tbl:fitparams}.  Though the
uncertainty is large in the photon-starved shock heated region, it is
clear that there is a significant increase in temperature in this
region compared to the surrounding regions.  
  
\subsubsection{Mach Number}
\label{subsubsec:m0744_mach}
We calculate the Mach number of the shock-front by fitting the
Rankine-Hugoniot jump conditions.  This quantity can be obtained
independently by fitting the jump in density from X-ray surface
brightness or in temperature as measured by spectroscopy.  We
use the analytic expressions from \citet{fino10} for the
Mach number in these two cases

\begin{equation}
\label{equation:mrho}
\mathcal{M}_{\rho}=\left[       \frac{ 2\frac{\rho_{2}}{\rho_{1}}}
{\gamma+1-(\gamma -1)\frac{\rho_2}{\rho_1}}        \right]^{1/2}
\end{equation}
and
\begin{equation}
\mathcal{M} _{T} =
\left\{\frac{ 8 \frac{T_2}{T_1} - 7 + \left[ \left( 8 \frac{T_2}{T_1} - 7 \right)^2 + 15 \right]^{1/2} }
{5}\right\}^{1/2} \, ,
\label{equation:mt}
\end{equation}
where we assume the adiabatic index for a monatomic gas
$\gamma=\frac{5}{3}$ and $\rho_{1}$, $\rho_{2} $, $T_{1}$ and $T_{2}$ are the
density and temperature before and after the shock.  

The Mach number can also be calculated from the stagnation
condition. This relates the ratio of the pressure at the edge of the
cold-front, $P_{st}$, over the pressure just ahead the shock-front,
$P_{1}$, to the Mach number through the relationship 
\begin{equation}
\label{equation:mstag}
\frac{P_{st}}{P_{1}} = \mathcal{M}_{st}^{2}\left(\frac{\gamma
    +1}{2}\right)^{\frac{\gamma +1}{\gamma -1}}\left(\gamma -
  \frac{\gamma -1}{2\mathcal{M}_{st}^{2}}\right)^{-\frac{1}{\gamma -1}}
\end{equation}
as presented in \citet{craigmergers}.

\begin{figure}
\plotone{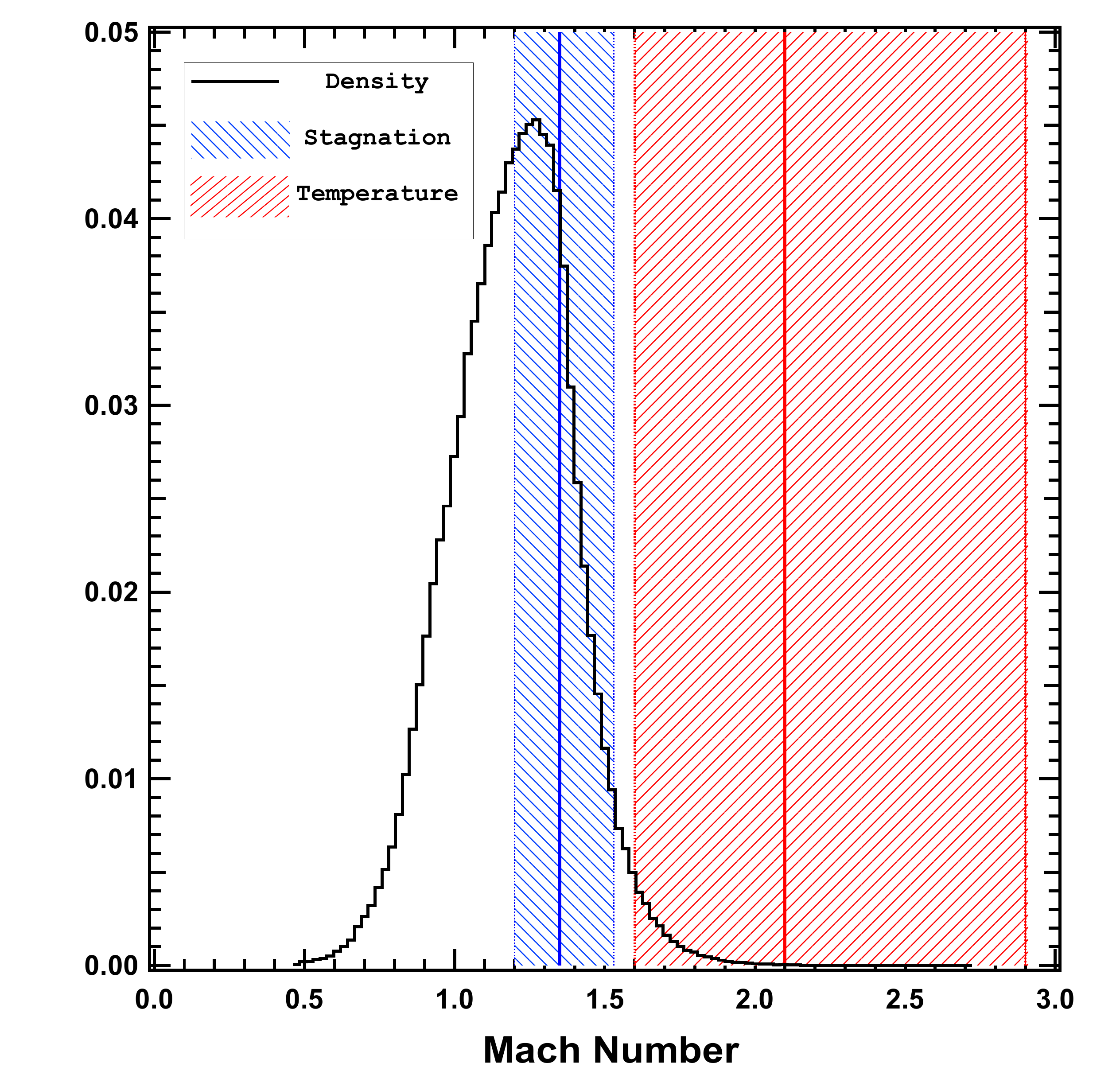}
\caption{Mach number in MACS0744 obtained with 3 methods. The black histogram shows
the discrete PDF calculated from the density jump conditions in
the MCMC fit to the surface brightness distribution.   The red
area shows the $68\%$ confidence region
from the calculation of a temperature jump across the shock-front as measured
from from \chandra\ spectroscopy. The blue area is the $68\%$ confidence region
obtained by fitting the stagnation condition.  The solid red and blue lines are
the best fit values obtained from the temperature jump and stagnation
conditions respectively.  While the stagnation and density jump
conditions yield highly consistent results, the result from the
temperature jump conditions appear to be biased high.  This is due to
a heavy reliance on the spectroscopy in the low SNR region II.}
\label{fig:machno}
\end{figure}

We calculate the Mach number for the potential merger in MACS0744 
using Equations \ref{equation:mrho}, \ref{equation:mt} and
\ref{equation:mstag}. The value obtained from the density jump
conditions was calculated from the posterior MCMC used in the fit
to the X-ray surface brightness. This yielded the value
$\mathcal{M}_{\rho}=1.2^{+0.2}_{-0.2}$ where the errors are $1\sigma$ and the
full discrete probability distribution function is shown in Figure
\ref{fig:machno}. 
The Mach number obtained from the relation imposed by the
stagnation condition is
$\mathcal{M}_{st}=1.4^{+0.2}_{-0.2}$ which is in excellent agreement with the
number provided by fitting the density jump.  
The temperature jump conditions at the shock yield a higher value, $\mathcal{M}_{T}=2.1^{+0.8}_{-0.5}$.  While this
measurement suggests a greater shock velocity, the error bars are
large and it agrees at the $1.3\sigma$ level with our estimate from
the density jump condition.
The flux scale in the
MUSTANG image suggests the true temperature is towards the low end of
the \chandra\ range as is shown
in Figure \ref{fig:m0744model}.  The shock
velocity in this cluster is $1827^{+267}_{-195}$ km s$^{-1}$ assuming
the Mach number obtained from the density jump conditions.

\subsubsection{Discussion}
\label{subsubsec:disc}

This high-redshift system has proved to be an excellent example of the
power of combining resolved SZE and X-ray imaging. 
The high-resolution SZE measurements reveal a region which is likely
the result of a shock.  Guided by this data, two sharp discontinuities
and a spectrum consistent with a substantially hotter plasma are
detected in the low SNR X-ray data.
Deeper \chandra\ observations of this cluster will help confirm the
presence of a shock and more accurately determine its Mach number,
which for the density jump fit to the current data is mildly
consistent with a transonic event ($\mathcal{M}=1$).

Figure \ref{figure:m0744composite} shows a composite image of this
system including the strong lensing mass distribution
\citep{richardlens}(see also \citet{jones10}).  This reveals a highly asymmetric
elliptical mass distribution elongated to the west consistent with
the  red sequence member galaxy distribution presented in
\citet{kartaltepe08}. \citet{zitrin10} have also done a mass reconstruction
and independently obtained a similar mass distribution.  The
HST data shown in green contains multiple bright red elliptical
galaxies with BCG-like characteristics.  Galaxy $G1$ is coincident with
the X-ray surface brightness peak and is assumed to be the BCG of the
main cluster.  Roughly one arcminute to the west of the X-ray center, the lensing mass
reveals a second peak containing the bright red galaxies
$G2$ and $G3$.   While
the baryon distribution is elongated in the direction of this
potential sub-cluster, there is no
X-ray peak associated with it.  This is likely explained by
ram-pressure stripping during passage of the sub-cluster
through the main core. Galaxy $G4$ is another massive cluster member
located west of the main peak.  This too has a significant dark
matter halo with no apparent baryonic peak.  The presence of multiple 
peaks in dark matter and galaxy density with no accompanying baryonic
mass is suggestive of a merger scenario in which a smaller cluster
has passed through the main core, stripping it of its baryons and
producing a shock wave in the ICM.  The geometry of the westerly
elongated multiply peaked dark matter
distribution is qualitatively suggestive of a merger scenario in which
the shock-heated gas identified
by MUSTANG could have been produced. However, an accurate interpretation of the
merger dynamics requires detailed modeling through hydrodynamical
simulations.

\subsection{RXJ1347-1145 (z = 0.45)}
The rich cluster RXJ1347-1145 is an extremely  X-ray luminous
galaxy cluster \citep{schindler97,allen02} and has been the
object of extensive study in SZE, X-ray, lensing, radio and optical
spectroscopy
\citep[e.g.,][]{schindler97,pointecouteau99,komatsu01,allen02,cohen02,kitayama04,Gitti07,ota08,bradac08,miranda08}.
It is one of only a few systems which have
been studied with the SZE on sub-arc minute scales thus far and is an excellent example
of the potential for the SZE to reveal the rich sub-structure
exhibited in
clusters.  Initial measurements made with ROSAT reported by
\citet{schindler97} deemed RXJ1347-1145 a relaxed system as it showed a round, singly
peaked surface brightness morphology and a strong cool core.
SZE observations made with the NOBA bolometer camera on the Nobeyama 45m
at 150~GHz
\citep{komatsu01,kitayama04} revealed an enhancement to the SZE  $20'' \, (170 \,
{\rm kpc})$ to the south-east of the cluster center.  This asymmetry
is now supported
by X-ray and radio data \citep{allen02,Gitti07} and is 
interpreted as a hot ($k_BT_e > 20 \, {\rm keV}$) feature caused by shock
heating in a recent merger event \citep{kitayama04}.  This feature was confirmed recently
by MUSTANG \citep{mason10}.  

\begin{figure}[tbh!]
\epsscale{1.5}
\plotone{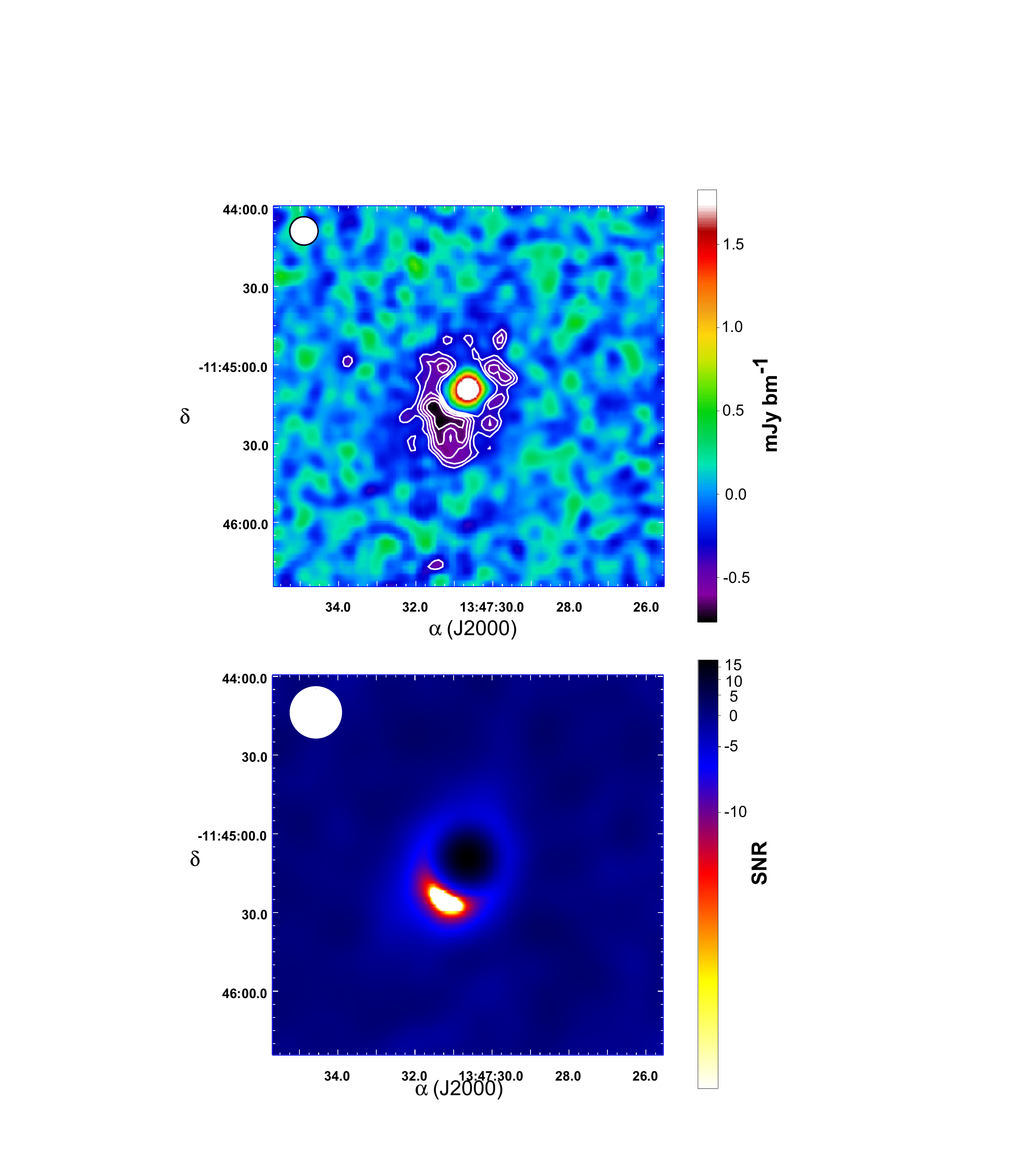}
\caption{Top: MUSTANG map of RXJ1347-1145 at $10''$ resolution.
  Contours are in units of $1\sigma$ starting at $3\sigma$.  Bottom:
  Signal to noise map convolved to $18''$ resolution.  The peak in the
  south-east quadrant, originally identified by Nobeyama at
  $4.2\sigma$ \citep{komatsu01}, is $13.9\sigma$.}
\label{fig:rxj}
\end{figure}

The data we present in
this paper are identical to those described in \citet{mason10}
but are processed with the additional per-pixel
high-pass Fourier filter described in Section \ref{sec:data}. 
By deliberately isolating the high frequency spatial modes we are able to
increase the signal to noise on the small-scale features shown in
Figure \ref{fig:rxj}.  
This comes at the expense of removing extended SZE signal.  While
the bulk emission is useful for measuring cluster parameters like gas
fraction and mass, MUSTANG's niche is constraining core sub-structure.
Similar techniques,
such as the subtraction of a $\beta$ model \citep{mcbubbles} or un-sharp
masking \citep{russel10} are
used in the X-ray to remove bulk emission and examine sub-structure.

The south-east
enhancement is likely due to shock-heated gas caused by a merger, but
the geometry and direction of propagation are not obvious as
in the case of MACS0744.  This makes it difficult to fit the
Rankine-Hugoniot jump conditions across a discontinuity in density inferred
from X-ray surface brightness. 
\citet{komatsu01} first
reported the south-east enhancement at a peak decrement of
$4.2\sigma$.  Figure \ref{fig:rxj} contains a signal to noise map
of the MUSTANG data set smoothed to comparable resolution to that of the
original Nobeyama map.  This result is now confirmed at a
$13.9\sigma$ significance level in SZE at $18''$ resolution.

We find good qualitative agreement of our map to a model cluster selected
from a suite of hydrodynamical simulations by \citet{sloshing}.  This
particular simulated cluster had recently undergone an off-axis merger
with a high mass ratio ($\sim10:1$).  At the epoch of observation, the sub-cluster is moving
through the atmosphere of the main cluster on its first pass of the
merger.  This model also reproduces several other
phenomenological features such as a sharp edge in X-ray surface
brightness to the east of the core
caused by sloshing of the cold gas, and the location of a second
bright elliptical galaxy, thought to be the BCG of the infalling
sub-cluster. 

\subsection{CL1226.9+3332 (z = 0.89)}
With a mass of ($1.4 \pm 0.2$)$\times 10
^{15}~\rm M_{\sun}$
\citep{jee09} within $r_{200}$, this system is among the largest known in the high
redshift Universe. Early measurements of the baryons were reported by \citet{ebeling1226} 
who identified it in the ROSAT WARPS survey. With the limited
resolution of ROSAT, the cluster 
was deemed to display relaxed morphology.  

Because of its high-redshift, X-ray spectroscopy on this object is 
difficult. Initial temperature measurements by
\chandra\
\citep{bonamente06} indicated a hot ICM ($\sim 14$
keV).  This was
consistent with \citet{maughan04} who made previous measurements with XMM ($\sim12$ keV).
A more detailed analysis of the ICM properties by \citet{maughan07} combined \chandra\ and XMM
spectroscopy. They confirmed the hot ICM and found an asymmetry in the
temperature map with the cluster emission south-west of the cluster center hotter
than ambient.  This object has also been mapped in the SZE on
arcminute scales by the SZA and a strong central decrement was measured
\citep{much07,tony09}. 

\citet{jee09} mapped the dark matter distribution of this system through a weak
lensing analysis.  They found that on large scales the cluster was
consistent with a relaxed morphology but the core was resolved into
two distinct peaks: the dominant one in close proximity to the BCG and
another $\sim 40''$ to the south west.
While this sub-clump shows no surface brightness peak in
either \chandra\ or XMM data, the location is consistent with the
temperature enhancement reported by \citet{maughan07} and a
secondary peak in the member galaxy density.  
One possible explanation of this is a merger scenario
in which a smaller cluster has passed through the dominant core on a
south west trajectory stripping its baryons and causing shock heating.

\begin{figure*}
\plotone{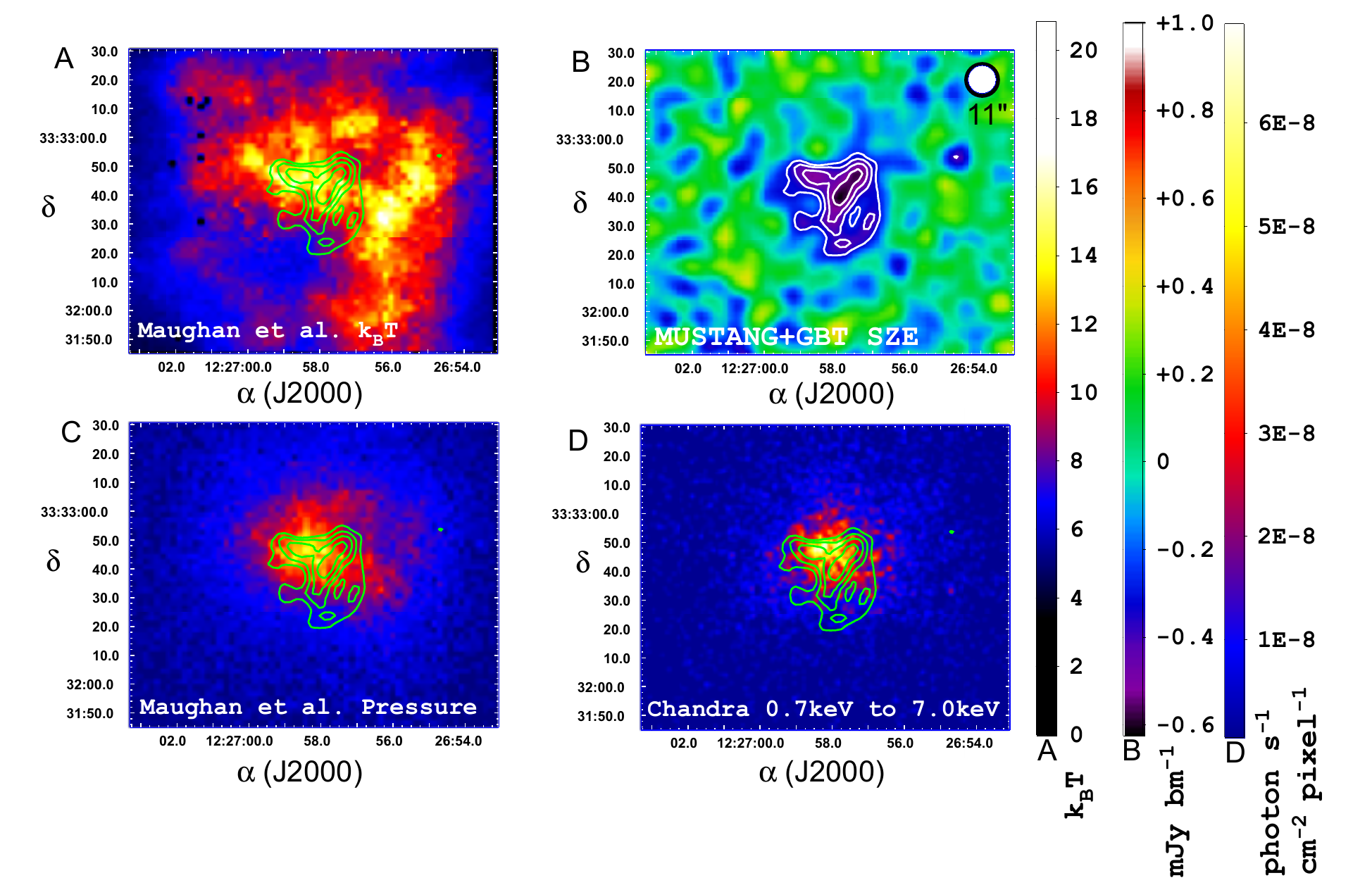}
\caption{Cl1226+3332 X-ray and SZE morphology. The contours on all
  images in this figure are MUSTANG SZE in units of $1\sigma$ starting
at $3\sigma$.  Panel $A$: Temperature distribution from
\citet{maughan07}.  Panel $B$: MUSTANG+GBT SZE image with $11''$ effective
resolution.  Panel $C$:  X-ray derived pseudopressure map from
\citet{maughan07}.  This was produced by taking the product of the
temperature map and the square root of the surface brightness. Panel $D$:
{\it Chandra} surface brightness in the $0.7$ keV to $7.0$ keV band smoothed
with a $1\farcs5$ Gaussian.}
\label{fig:four1226}
\end{figure*}

The MUSTANG map is shown in Figure \ref{fig:four1226}. 
It reveals an asymmetric,
multiply peaked pressure morphology in this high-$z$ system.  The most
pronounced feature is a narrow ridge $\sim20''$ long located
$\sim10''$ south-west of the X-ray peak. A second peak is found in
good proximity to the X-ray emission which is also coincident with the
BCG.   Also shown in this figure is the
X-ray derived temperature and pseudopressure (defined as the product
of the temperature map and the square root of surface brightness) maps from
\citet{maughan07}. The \chandra\ surface brightness image was produced
with 74~ksec of archival data taken in ObsIds 3180, 5014 and 932.
There is good qualitative agreement between the
two data sets, although the small
scale features seen by MUSTANG are absent in the Maughan map.  This is
not unexpected as the X-ray derived pressure relies heavily on the
temperature map which was produced with a variable sized
aperture.  Therefore, adjacent pixels are not independent.  This correlation
makes the map less sensitive to small-scale features.

\begin{figure}
\plotone{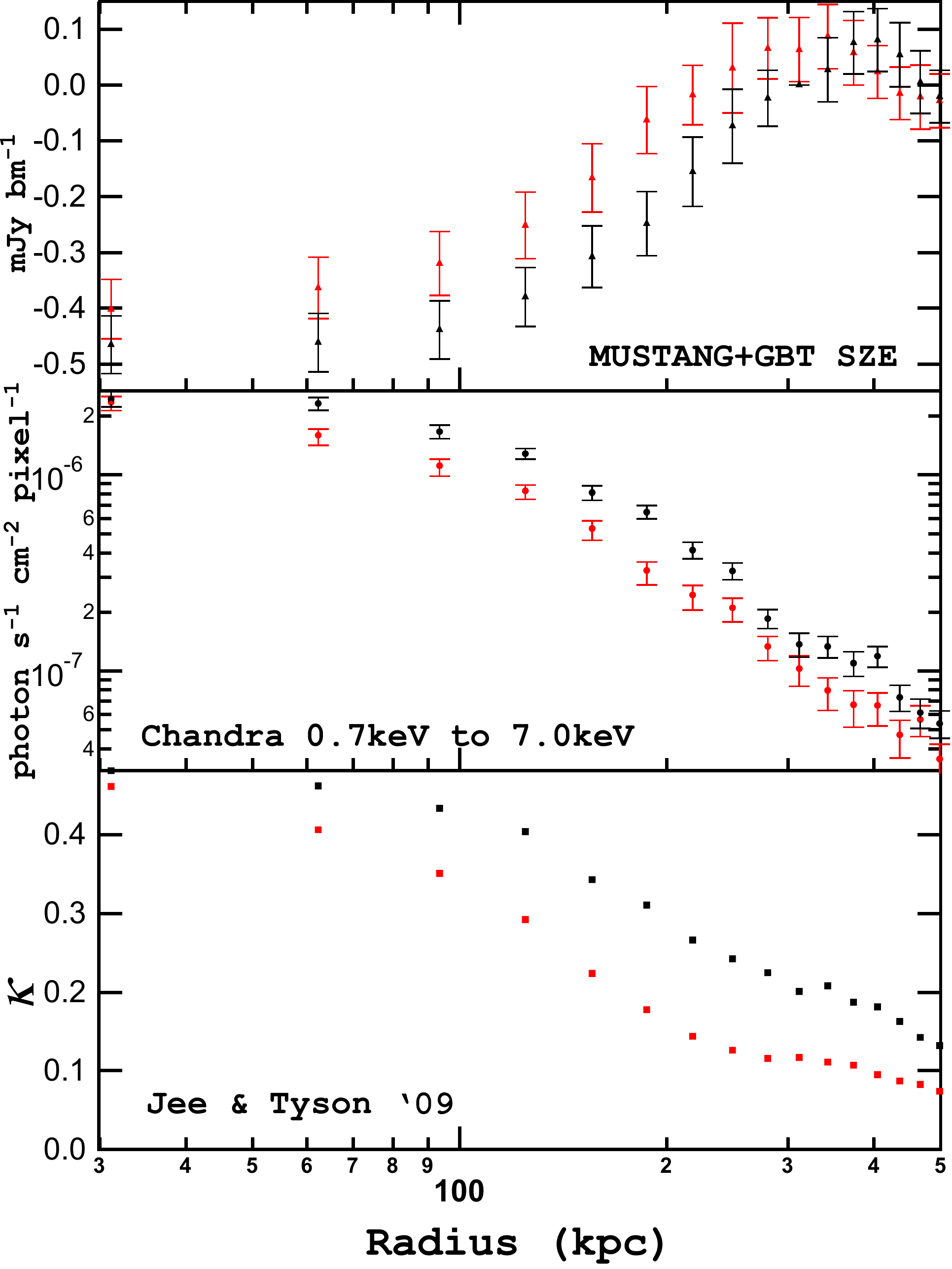}
\caption{Radial profiles of CL1226+3332 from the SZE (top), X-ray surface brightness
  (middle) and lensing mass distribution from \citet{jee09}(bottom).
  Profiles are centered on the X-ray peak and are taken from the
  south-eastern quadrant (red) and south-western quadrant(black). The
  SZE map was convolved with a $10''$ Gaussian before averaging. All
  data sets are consistent with an elongation in the south-west
  direction as proposed by the merger scenario}
\label{fig:profs1226}
\end{figure}

Figure \ref{fig:profs1226} shows radial profiles of the X-ray surface
brightness, SZE and lensing mass distribution.  The profiles are
centered on the X-ray peak which is coincident with the BCG.  Each
plot shows two profiles, one taken from 
the south-eastern quadrant (red) and the other in the south-western
quadrant (black).  It is clear that all data sets are consistent with
an asymmetry elongated towards the south-west as proposed by the
merger scenario. 

\begin{figure}
\plotone{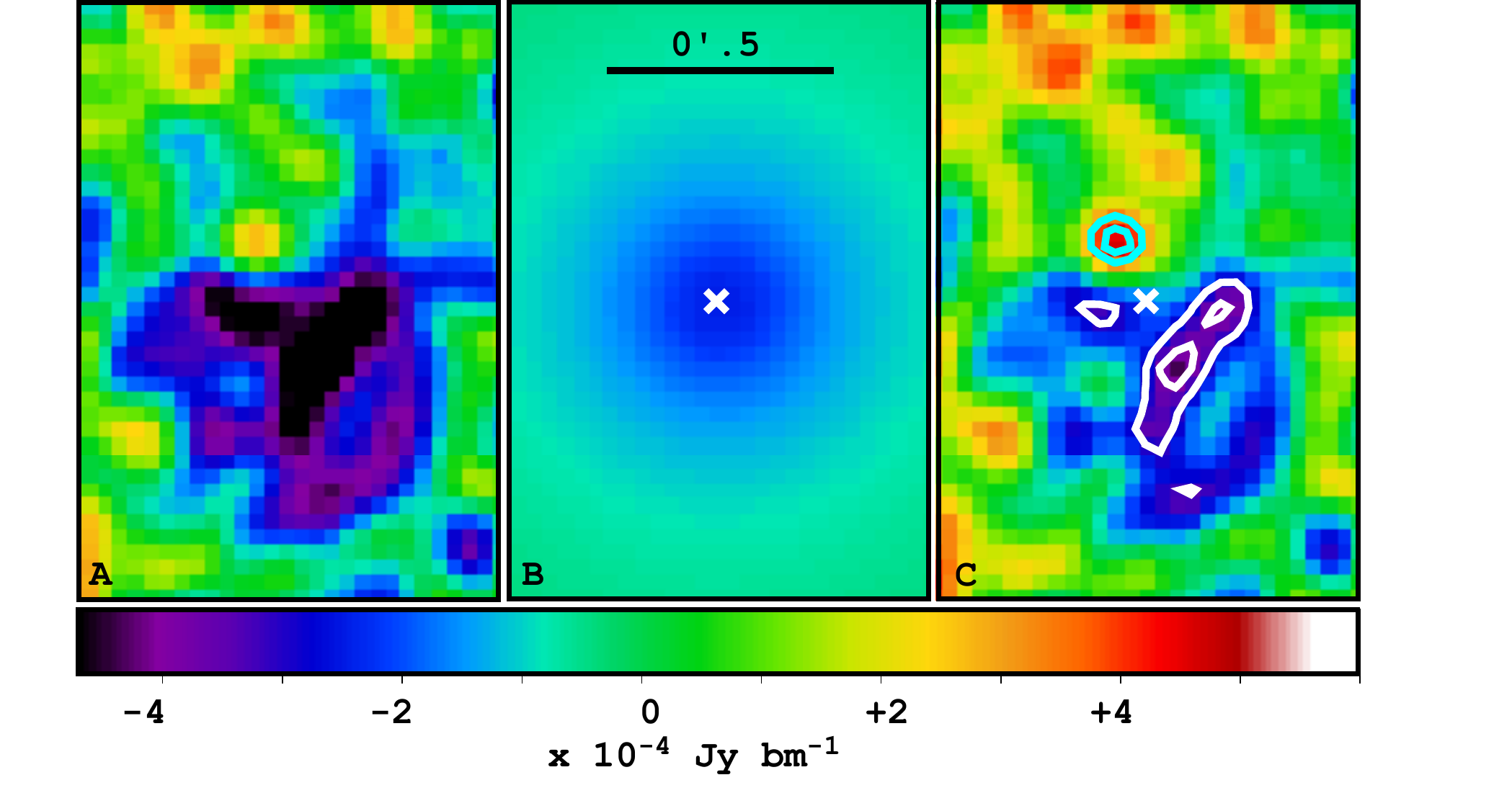}
\caption{Panel A:  MUSTANG SZE map of the core in CL1226.  Panel B:
  The best fit model of a \citet{nagai07} profile to SZA data as is
  presented in \citet{tony09}.  The model has been passed through the
  appropriate transfer function.  Panel C: The residual of Panel A -
  Panel B. The white contours are ($-3\sigma$,$-4\sigma$) which show
  the significance of the sub-structure not accounted for in the
  azimuthally symmetric model.  The cyan
  contours are ($+3\sigma$,$+4\sigma$).  The white x shows the location of the
  X-ray peak as measured by {\it Chandra}.}
\label{fig:sub1226}
\end{figure}

The core of this cluster is compact on the sky due to its high
redshift.  For this reason, the bulk emission is likely to contribute
non-negligible amounts of flux to the MUSTANG map.  To quantify the
significance of the sub-structure, we compare our map to the best fit
spherically symmetric \citet{nagai07} model of the SZA data as presented by \citet{tony09}.
Figure \ref{fig:sub1226} shows our map.  We assume the spherically
symmetric bulk model is centered on the X-ray peak and take the
difference between the two maps.  The residual map
figure, contains the peak of the ridge at a $4.6\sigma$ level.

There is a positive unresolved feature in the residual map in Figure
\ref{fig:sub1226} located $8''$ northwest of the X-ray peak.
This unexplained feature could have several interpretations.
It could simply be a noise artifact.  However, it is also possible
that it is a faint unresolved source.  Because it is not detected at
30~GHz \citep{tony09}, such a source would require a rising spectrum in
the millimeter as would be expected from a high-redshift, dusty
star-forming galaxy.  This galaxy could be lensed as speculated by
\citet{blain2002} and \citet{lima2010} and similar to the one found in
the Bullet cluster by \citet{wilson2008} and \citet{rex2009}. 
Disentangling speculation such as
this requires the addition of resolved millimeter or submillimeter
follow-up with different instruments.

\begin{figure*}
\plotone{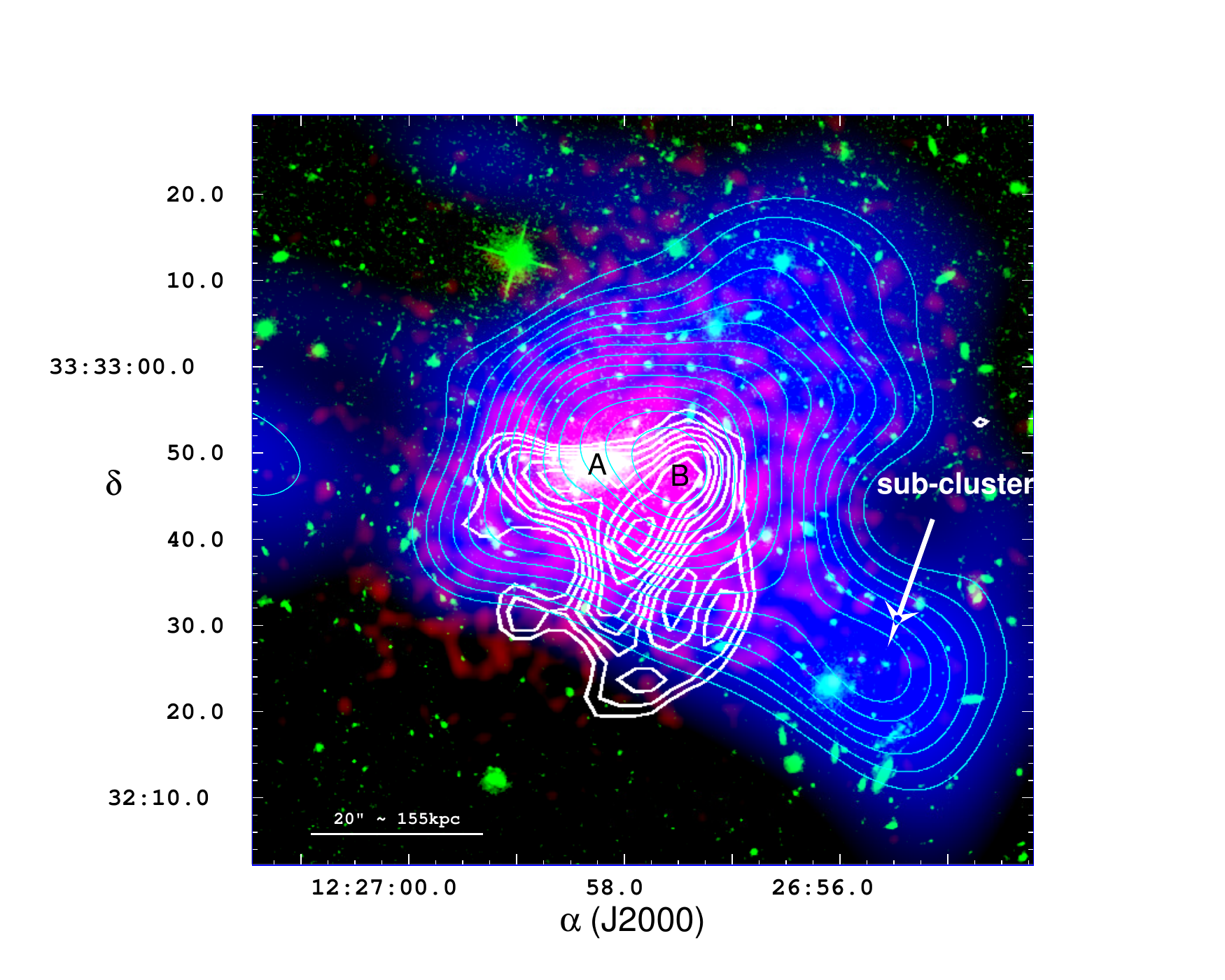}
\caption{Composite image of CL1226.  Red shows the \chandra\ surface
  brightness in the 0.7keV to 7.0keV band.  Blue color scale and cyan
  contours show the surface mass density distribution of
  \citet{jee09}.   Contours are linearly spaced in 15 intervals between $\kappa=0.25$
  and $\kappa=0.59$.  Green traces the optical emission as measured by
  the HST/ACS in the F814W band. White contours show the MUSTANG
  measurement in units of $0.5\sigma$ starting at $3\sigma$. Location
  $A$ demarcates the BCG and is 
  coincident with the X-ray surface Brightness peak.  Location $B$ shows
  the Dark Matter peak which is coincident with the northern lobe of the
  SZ ridge revealed by MUSTANG imaging.}
\label{fig:composite1226}
\end{figure*}

A multiwavelength composite image of this system is presented in
Figure \ref{fig:composite1226} which includes the weak lensing mass
distribution presented in \citet{jee09}.  The northern end of the dominant ridge in the
MUSTANG image, labelled ``B'' in this figure, is roughly coincident with
the lensing mass peak.  The orientation of the ridge is approximately
orthogonal to a vector connecting the BCG and secondary lensing
peak which is most likely the trajectory of the sub-cluster.  We posit
that the ridge is produced by  a reservoir of shock-heated gas
created in the core passage of the sub-cluster, reminiscent of the
eastern peak in the famous ``Bullet Cluster'' \citep{mark02}.  

\subsection{Abell 1835 (z = 0.25)}
The massive cool core cluster Abell 1835 has proved to be an excellent
laboratory for studying a range of cluster physics.  It
has been used to map the large scale dark matter distribution
\citep{lensing1835}, look for effects of small-scale turbulence
\citep{turbulence1835}, search for lensed background submm galaxies
\citep{submm1835}, map the extended radio emission \citep{radio1835}
and study the central cool core
\citep[e.g.,][]{peterson1835,schmidt1835}. This cluster has also been the
subject of extensive SZE modeling
\citep{reese02,benson04,bonamente06,bonamente2008,bolocam1835}.

\begin{figure*}
\plotone{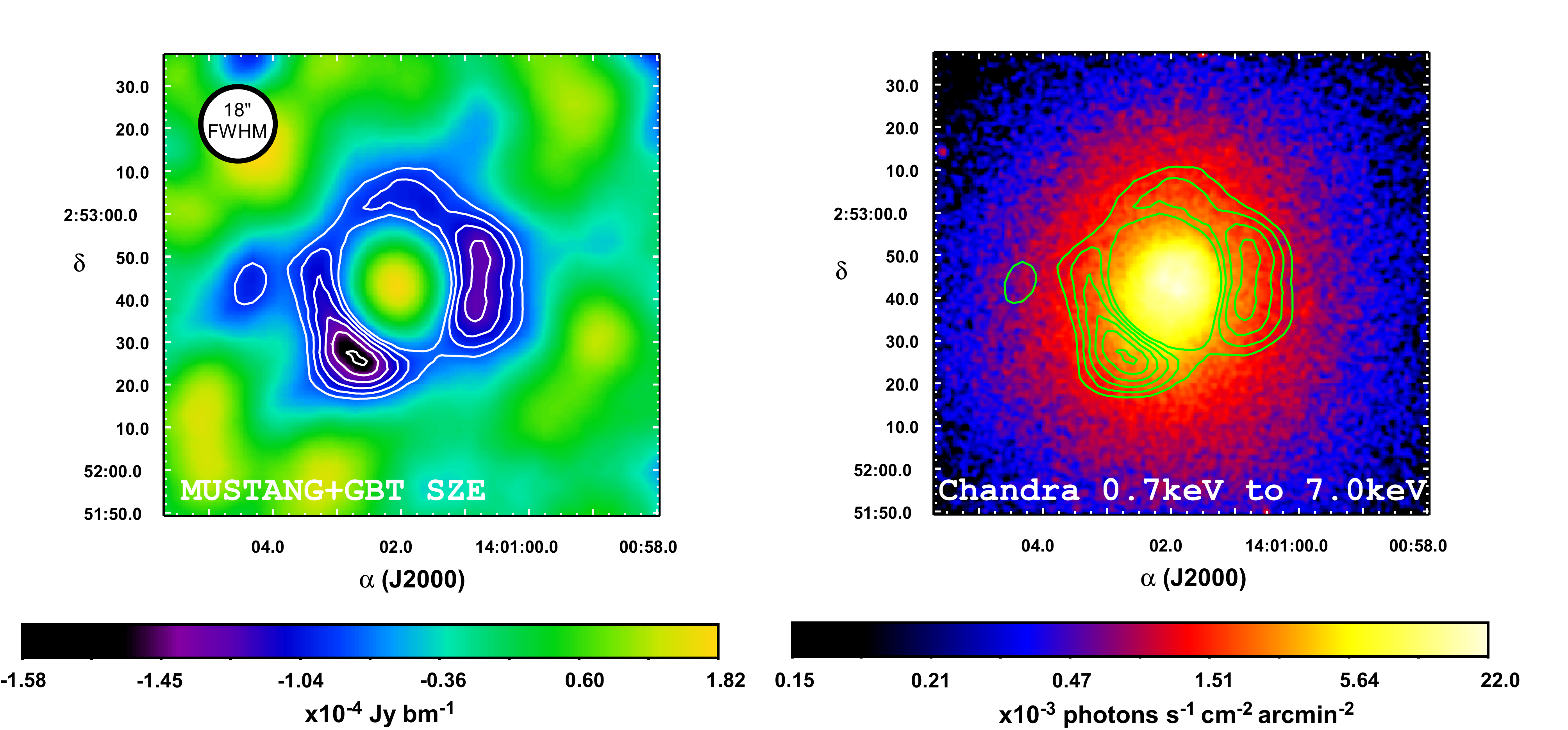}
\caption{Left: MUSTANG SZE image of Abell 1835 smoothed to $18''$
  resolution.  Contours are units of $0.5\sigma$ starting at
  $2.5\sigma$.  Note the central unresolved radio source. Right: {\it
    Chandra} $0.7$ keV to $7.0$ keV image of Abell 
  1835 smoothed with a $1.5''$ Gaussian.  Contours on the right are
  identical to those on the left.}
\label{fig:a1835}
\end{figure*}

Aside from the the central $\sim10''$ region which displays a cavity system excavated by a central AGN
\citep{mcnamara1835_bubbles}, the X-ray morphology  is well
described by a spherically symmetric geometry with no
obvious sub-structure.  This distinguishes it from the rest of our
sample.  The absolute pressure is extremely high in the core, as was
demonstrated by \citet{reese02} who
measured a central decrement of $-2.502^{+.150}_{-.175}$ mK at 30GHz.  However,
the MUSTANG map shown in Figure \ref{fig:a1835} contains a low
signal to noise detection of the SZE.  The map in this figure has been
smoothed to an effective resolution of $18''$ to increase the signal
to noise.  This figure also shows the \chandra\ image produced from
222 ksec of data, merged from ObsIds 495, 496, 6880, 6881 and 7370.
As described in Section \ref{sec:data}, the filtering
techniques applied are optimized to produce high signal to noise maps
on small-scale structures. Therefore, the lack of high significance SZE in
the reconstructed image is indicative of a featureless, smooth, broad
signal.

\begin{figure}
\plotone{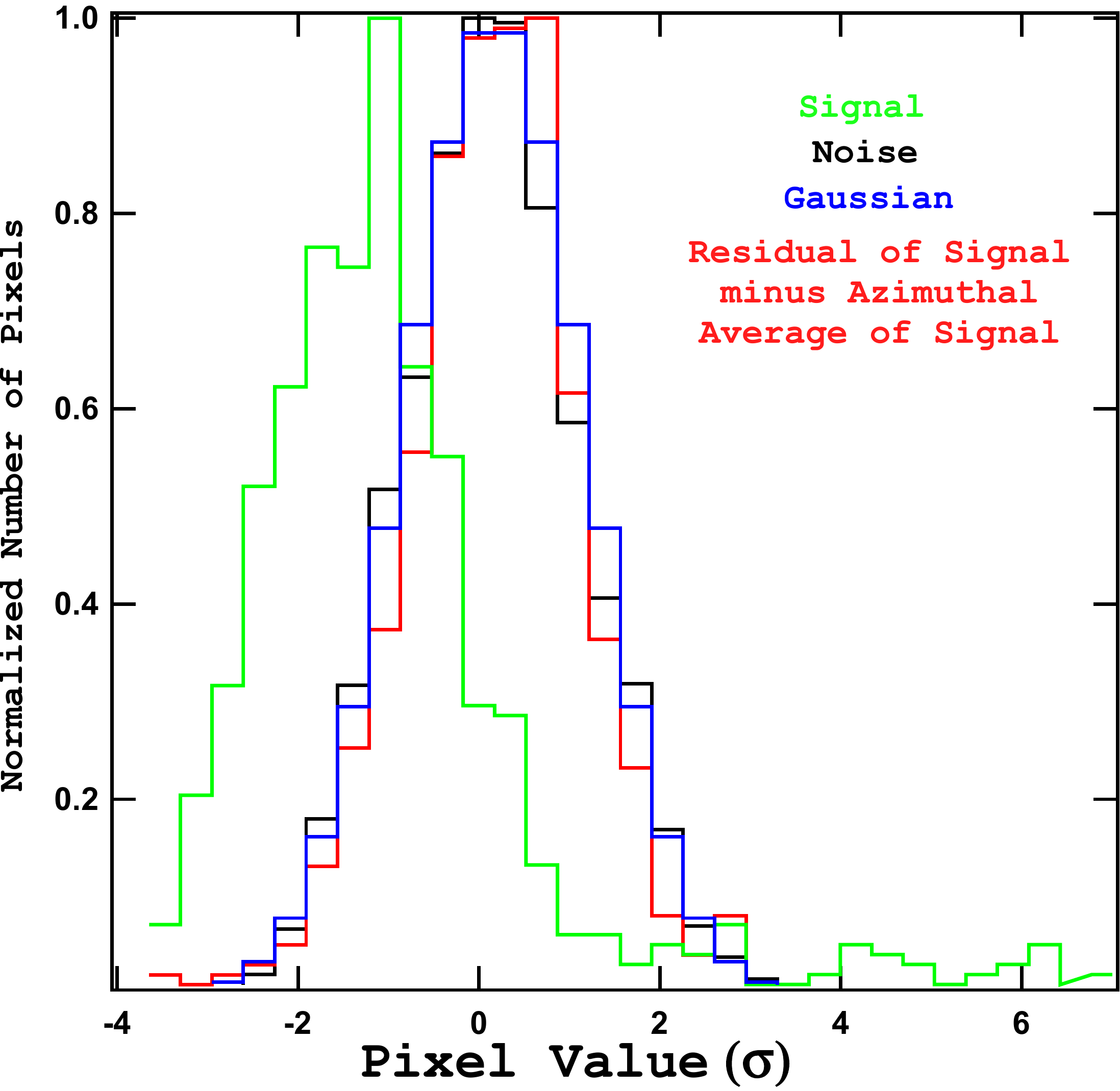}
\caption{Pixel histograms showing the azimuthal symmetry of the
  MUSTANG map in Abell1835.  To account for uneven noise distributions
caused by non-uniform exposure, histograms are from pixels in the map
multiplied by the square root of the weight map.  Green is the
histogram of the central $1'$ diameter with the signal in it.  The map
used was smoothed to $11''$ effective resolution.  Note the positive tail
caused by the central point source.  While the peak in SNR
of the SZE is moderate in a single beam, the area used for this
histogram is larger than 20 beams.  The black histogram is from an
area off source believed to contain no signal.  Blue is a Gaussian
distribution.  Red shows a histogram over the same area as the green
line after the subtraction of an azimuthal average of the data.  Units
of the abscissa are multiples of $1\sigma$ of the
Gaussian distribution.  To account for the different number of pixels
in the two regions, all histograms are normalized to peak at 1.  From
this analysis we conclude that the MUSTANG
signal is consistent with containing no significant structure that does not have azimuthal
symmetry.}
\label{fig:hist1835}
\end{figure}

Figure \ref{fig:hist1835} shows several pixel histograms taken from
different areas in the MUSTANG Abell 1835 map.  To account for uneven
exposure, the maps were multiplied by the square root of the weight
map.  This figure contains a pixel histogram of the central $1'$
diameter in the MUSTANG map minus
an azimuthally averaged version of the same map. This is shown alongside
histograms from a region off-source assumed to contain negligible
signal and a Gaussian distribution with its $\sigma$ equal to the
standard deviation of the pixels in the off-source region.  The
residual of the data and azimuthally averaged data shows no
significant deviation from the noise map and is well described by a
Gaussian.  From this, we conclude that the data are consistent with
containing negligible signal which deviates from spherical symmetry.

\section{SZE Flux estimates}

We provide in this section estimates of the SZE flux measured by
MUSTANG.  We quote the integrated flux within two radii,
$\theta_{2\sigma}$ and  $\theta_{3\sigma}$, which correspond to the
radii where the mean
significance is 
$>$2$\sigma$ per beam, and again at $>$3$\sigma$ per beam,
binned over many beams.
We compare our estimates to extrapolations of the SZE flux
reported  by \citet{bonamente2008}, which was computed within $r_{2500}$ using 
100~kpc core-cut $\beta$-model fits to the lower resolution 30 GHz OVRO/BIMA data sets.  
We use the \citet{itoh1998} relativistic corrections to the SZE flux frequency
relation (described in Appendix \ref{section:etherm}) when scaling to 90 GHz, assuming the isothermal temperatures 
reported by \citet{bonamente2008}.
Any bias due to non-isothermality as well as any discrepancy between our temperatures and those 
reported in \citet{bonamente2008} leads to $<2\%$ bias in this rescaling, which is well within
the calibration and compact radio source contamination uncertainties in both measurements.

For A1835, CL1226, and RXJ1347, we bin the flux per pixel within a circular region 
centered on the peak SNR of the map.
For MACS0744, we choose an elliptical region to capture, approximately, 
the shape of the prominent SZE sub-structure (i.e. the shock-heated region reported in 
Section \ref{subsubsec:m0744_mustang}). 
Uncertainties in the absolute calibration are on the order of $\lesssim15\%$, which
we include in our estimates.
In order to account for any compact radio sources in these regions, we extrapolate the
flux measurements from FIRST \citep{first} and NVSS \citep{nvss} at 1.4~GHz and SZA, OVRO, and BIMA at $\approx 30$~GHz 
\citep{laroque06,tony09} to 90~GHz (see Table \ref{tbl:ptsrcs}) using
a power law fit, where the uncertainties are treated in quadrature.
The predicted fluxes of the radio sources are given in Table
\ref{tbl:ptsrcs}.  Variability in source flux is not accounted for although
the multi-band measurements were obtained at different
epochs.  We are forced to make the assumption of flux stability as
there are insufficient available time sampled radio data to do otherwise. 
Since we measure SZE flux as a decrement, we add these radio source flux estimates 
to our measurements to obtain estimates of the underlying SZE flux reported in
Table \ref{table:clusterflux}.  In the flux estimates presented, we
have accounted for attenuation by the filtering by dividing by the
mean amplitude of the angular transfer function over the Fourier modes
between the beam scale and the radius to which we are
integrating.  This value is typically $\sim0.7$.

SZE flux provides an estimate, without relying on X-ray data,
of the thermal energy in the ICM structure we see.  
Table \ref{table:clusterflux} contains the integrated flux and thermal
energy estimates for these four objects calculated using the methods
described in Appendix \ref{section:etherm}.
We include flux estimates at 30~GHz that were provided by fits to the SZA, OVRO,
and BIMA data \citep{laroque06,tony09}.   
It is important to note that these flux and energy estimates reported here only represent
structure on scales $\lesssim 30 \arcsec$, which remain after the 
optimal filtering described in Section \ref{sec:data}.  Contribution to the MUSTANG flux from the extended
bulk cluster
signal will be included in this number as well as the two are
degenerate.
Since residual emission from the
bulk has not been subtracted, these values should be regarded as upper
limits of the energy contained in the small-scale structure.

While Abell 1835 and RXJ1347 have comparable integrated flux on
large scales, MUSTANG measures a dramatic difference in the cores.  This
is caused by the large amount of signal on small-scales in the
sub-structure of RXJ1347 and the smooth featureless distribution of
Abell 1835.

\begin{deluxetable*}{lccccc}
\tablecaption{Unresolved Radio Sources}
\tablehead{
Cluster Field & \multicolumn{2}{c}{\underline{Coordinates (J2000)}}\footnotemark[1]    & Flux(1.4 GHz)  &  Flux(30 GHz)\footnotemark[2]  &  Flux(90 GHz)\footnotemark[3] \\ 
        & \multicolumn{2}{c}{$\alpha$~~~~~~~~~~~~~~~~~~$\delta$} & NVSS/FIRST(mJy) & (mJy) & (mJy)} 
\startdata
Abell~1835        & $14^h 01^m 02^s\!.1$ & $+02^{\circ}52^{\prime}43\arcsec\!.2$ & 31.25$\pm 1.57$/39.32$\pm 1.56$ & ~2.8$\pm0.3$  & 1.2$\pm0.2$/1.1$\pm0.2$\\
RX~J1347.5$-$1145 & $13^h 47^m 30^s\!.7$ & $-11^{\circ}45^{\prime}08\arcsec\!.6$ & 45.89$\pm 1.46$/NA              & 10.4$\pm0.3$  & 6.2$\pm0.3$ \\
RX~J1347.5$-$1145 & $13^h 47^m 30^s\!.1$ & $-11^{\circ}45^{\prime}30\arcsec\!.2$ & 17.66$\pm 3.16$/NA              & $<$0.3        & $<$0.07 \\
CL~J1226.9+3332   & $12^h 26^m 58^s\!.2$ & $+33^{\circ}32^{\prime}48\arcsec\!.6$ & ~3.61$\pm 0.22$/4.34$\pm 0.47$  & $<$0.2        & $<$0.13 \\
\enddata
\footnotetext[1]{Coordinates are from FIRST except in RXJ1347 where they come from NVSS.}
\footnotetext[2]{Measured by OVRO, BIMA and the SZA.}
\footnotetext[3]{Extrapolated assuming a power law spectral energy distribution.}
\label{tbl:ptsrcs}
\end{deluxetable*}
\begin{deluxetable*}{llcccccccc}
\label{table:clusterflux}
\tablecaption{Cluster Flux Estimates From MUSTANG}
\tablehead{
Cluster Name      & $z_r$   & $D_A$    & $\theta_{2500}$      & $Y$\footnotemark[1]                      & $|F_{\rm SZE}(90~\rm GHz)|$\footnotemark[2]  & $|F_{\rm SZE,MUSTANG}|$\footnotemark[3]      & $\theta_{3{\sigma}}$ & $\theta_{2{\sigma}}$ & $E_{th}$ \\ 
                  &     & (Gpc)    & ($\arcsec$)          & (10$^{-10}$)             & (mJy)                               & (mJy)                       & ($\arcsec$)           & ($\arcsec$)           & ($10^{62}$ erg) }\\
\startdata
Abell 1835        &0.25 & $ 0.81 $ & $172\pm^{  5}_{  4}$ & $2.09\pm^{0.17}_{0.16}$  & $174\pm^{14}_{13}$                  &  2.4--3.4$\pm^{0.3}_{0.3}$  & 22.9                  & 27.3                  & 0.6--0.9\\[.25pc]
RX~J1347.5-1145   &0.45 & $ 1.19 $ & $122\pm^{  4}_{  4}$ & $1.62\pm^{0.18}_{0.18}$  & $135\pm^{15}_{15}$                  & 12.9--18.5$\pm^{2.3}_{2.3}$ & 18.8                  & 22.1                  & 7.1--10.2\\[.25pc] 
MACS~J0744.8+3927 &0.69 & $ 1.47 $ & ~$59\pm^{  3}_{  3}$ & $0.34\pm^{0.04}_{0.04}$  & ~$28\pm^{3.3}_{3.3}$                &  0.8--1.2$\pm^{0.1}_{0.1}$  & ~6.8                  & ~9.8                  & 0.7--1.0\\[.25pc]
CL~J1226.9+3332   &0.89 & $ 1.60 $ & ~$66\pm^{  7}_{  6}$ & $0.35\pm^{0.05}_{0.05}$  & ~$29\pm^{4.2}_{4.2}$                &  2.1--2.6$\pm^{0.4}_{0.4}$  & 15.0                  & 17.9                  & 2.1--2.6\\ 
\enddata
\label{table:clusterflux}
\footnotetext[1]{$Y=\int y_{C}d\Omega$.}
\footnotetext[2]{From model fits to OVRO/BIMA by \citet{bonamente2008} at 30~GHz scaled to 90~GHz.}
\footnotetext[3]{First and second numbers correspond to totals from all pixels within radii where the mean significance is greater than 3$\sigma$ ($\theta_{3{\sigma}}$) and 2$\sigma$ ($\theta_{2{\sigma}}$) per beam respectively.}
\end{deluxetable*}



\section{Conclusions}
In this paper we have presented high-resolution images of the
Sunyaev-Zel'dovich effect in four massive galaxy clusters produced
from MUSTANG observations. 
Three of the four systems probed here display sub-structure
in the SZE.  In the case of MACS0744, we identify a likely shock-front
propagating with a Mach number of $\mathcal{M}=1.2^{+0.2}_{-0.2}$.  The 
shock-heated kidney shaped
feature is located between the 
system's main mass peak and a
second peak which shows no evidence of significant baryonic mass. In our highest
redshift system, CL1226, we find a multiply peaked pressure
distribution with an asymmetric morphology.  The location and
orientation of a ridge found in the SZE, along with a south-westerly
elongated shape, are qualitatively supportive of the merger
scenario proposed by \citet{jee09}.  We also present a new reduction of the
data from observations of RXJ1347 presented in \citet{mason10}.  This higher signal to
noise map confirms the previously reported south-east pressure
enhancement at a $13.9\sigma$ confidence level. In Abell 1835 we report a
detection consistent with a spherically symmetric pressure
distribution and no significant sub-structure.

This pilot study has demonstrated the potential of high-resolution SZE
to identify sub-structures such as weak shocks in galaxy clusters.  This
is particularly true of the high-redshift universe where
the X-ray data are photon starved. A next generation feedhorn-coupled
TES bolometer array for the GBT is currently in the planning stages.  With a much
larger FOV ($4\farcm5$) and a mapping speed 1000 times that of MUSTANG it
will be able to image a large number of clusters
on angular scales from $9''$ to $9'$.
Other instruments coming
on line in the next decade, such as ALMA, the LMT, SCUBA2 and CCAT will also
have high-resolution SZE capabilities.

\begin{acknowledgements}
The National Radio Astronomy Observatory is a facility of the National
Science Foundation operated under cooperative agreement by Associated
Universities, Inc. The observations presented here were
obtained with telescope time allocated under NRAO proposal ids AGBT08A056,
AGBT09A052, AGBT09C059 and AGBT10A056.
We would like to thank James Jee and Anthony Tyson for
providing their lensing map for CL1226, Johan Richard and J.P. Kneib 
for their lensing map in MACS0744, John Zuhone for providing a Compton
$y_C$ map
from his numerical simulation of RXJ1347, and Ben Maughan for his temperature
and pressure maps in CL1226. 
The contributions of Dominic Benford, Harvey Moseley, Johannes
Staguhn, Jay Chervenak, Kent Irwin, Peter Ade, Carole Tucker, Bill
Cotton and Mark Whitehead
were essential to the functionality of the instument.
The late night assistance of the GBT operators was much
appreciated during the observations.  We also thank James Aguirre,
Danny Jacobs and
Gary Bernstein for useful conversations.  Much of the work presented here
was supported by NSF grant AST-0607654.  PMK was also
funded by the NRAO graduate student support program. TM
is supported as a NASA Einstein fellow and obtained funding through
grant PF0-110077. CLS and MS were supported in part by \chandra\
grants G07-8129X, GO8-9083X, GO9-0135X and G09-0148X and XMM grants
NNX08AZ34G, NNX08AW83G and NNX09AQ01G. 
\end{acknowledgements}






\appendix
\section{A: Shock Model}
\label{section:shock}

\subsection{A.1 : Surface Brightness Profiles}

\begin{figure}[tbh!]
\epsscale{0.5}
\plotone{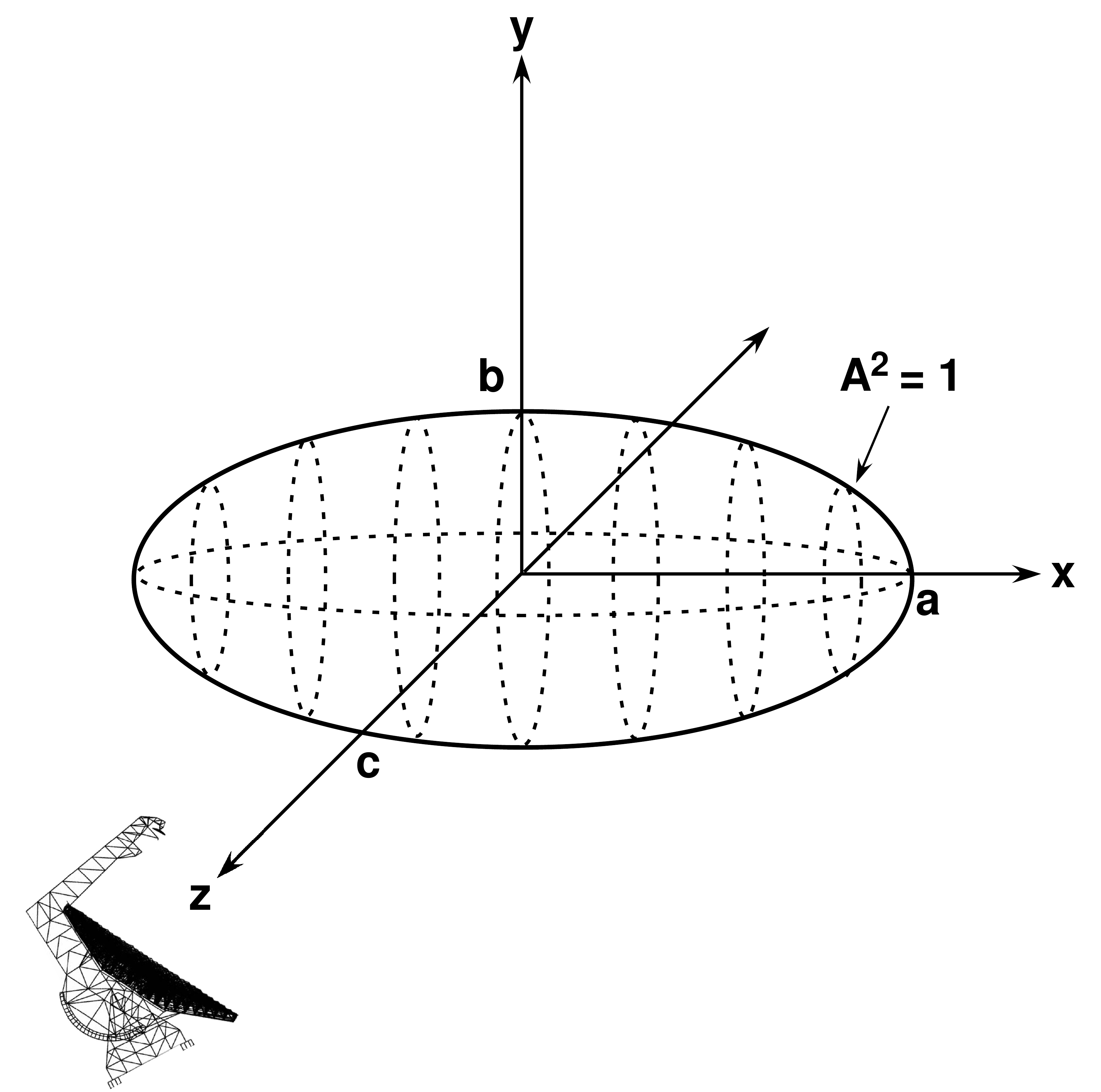}
\caption{Elliptical geometry for a single surface brightness edge used
in modeling shock fronts.}
\label{figure:ellipse}
\end{figure}

In this work, we measure the density characteristics of a shock-front and cold front in MACS0744 by
analyzing the elliptical profiles of X-ray surface brightness $I(x,y)$ in some observed photon energy band
$E_1$ to $E_2$.
(In this paper, we consider the surface brightness in the 0.7--7 keV band.)
Here, we give the analytic expressions for the X-ray surface brightness of elliptical X-ray images with discontinuities.
The X-ray surface brightness is given by the line of sight integral
\begin{equation}
\label{equation:mainint}
I(x,y)= \frac{1}{4\pi(1+z_{r})^{\eta}}\int \varepsilon(x,y,z ) \, dz \, ,
\end{equation}
where $\varepsilon$ is the X-ray emissivity integrated over all directions
in the emitted energy band $E_1 ( 1 + z_r)$ to $E_2 ( 1 + z_r)$.
The Cartesian coordinates $x$, $y$, and $z$ are aligned as shown in Figure~\ref{figure:ellipse}, with
$z$ being along the line of sight.
The cluster redshift is $z_{r}$.
The parameter $\eta$ is 4 if $I(x,y)$ is given in energy units, and $\eta = 3$ if $I(x,y)$ is in counts units, which is generally the case for X-ray observations.

We fit the data with an analytic expression for the above integral
obtained with the following assumptions:

\begin{enumerate}

\item The X-ray emissivity
  $\varepsilon (x,y,z)$ is constant on concentric, aligned, similar
  ellipsoidal surfaces with the geometry and conventions described in
  figure \ref{figure:ellipse}.
  The three principal axes of this elliptical distribution are $a$, $b$, and $c$.

\item Two of the principal axes of the distribution ($a$ and $b$) lie in the plane of
  the sky, and the third axis ($c$) lies along the line of sight.  We take the $x$ axis of our coordinate system to
  be parallel to $a$, and the $y$ axis to be parallel to $b$.
  The axis given by $a$ or $x$ is along the direction of propagation of the shock and/or cold front.

\item Between each of the discontinuities, the emissivity varies as a power-law of the radius, $\varepsilon = \varepsilon_{o} r^{-p}$. 
  Here, $r$ is the scaled elliptical radius
  $ r = [ ( x / a )^2 + ( y / b)^2 + ( z / c )^2 ]^{1/2}$
  and $p$ is the
  power law index.
  The emissivity changes discontinuously at the shock front and/or cold front.

\item The shock front and/or cold front has rotational symmetry about an axis in the
  plane of the sky along its direction of propagation ($c=b$).
  Although we make this assumption in our analysis of the data on MACS0744, none of the expressions
  given below depend on this assumption, and are correct for any $c$.

\end{enumerate}


We treat separately each of the regions bounded by one or two discontinuities.
In the case of a shock and cold front, there are three separate regions: 
the pre-shock gas, the shock-heated gas, and the cold front gas.
Since equation~(\ref{equation:mainint}) is linear in $\varepsilon$, we can then sum the surface brightnesses
of these regions to give the total surface brightness.

Since the plane of the sky corresponds to  a plane of symmetry at $z = 0$ in this model, the integral for the surface brightness can limited to positive $z$ and doubled, giving
\begin{equation}
\label{equation:mainint2}
I(x,y)= \frac{1}{2\pi(1+z_{r})^{\eta}}\int_{q_1}^{q_2} \varepsilon(x,y,z)\,dz \, .
\end{equation}
The values of $q_1 \ge 0$ and $q_2 \ge 0$ give the extent of the cluster region along the line of sight.
The general form for the surface brightness for each of the regions obtained with these assumptions after integration is
\begin{equation}
\label{equation:model}
I(x,y)=\frac{1}{4 \pi^{1/2} (1 + z_r)^\eta} \epsilon_{0}c\frac{\Gamma(p-\frac{1}{2})}{\Gamma(p)}A^{-2p+1}\phi
\, ,
\end{equation}
where we define $A$ to be the two-dimensional scaled elliptical radius
\begin{equation}
A(x,y) \equiv  \left( \frac{x^{2}}{a^{2}}+\frac{y^{2}}{b^{2}} \right)^{1/2} \, ,
\end{equation}
and $\Gamma$ is the standard Gamma function.
The piecewise function $\phi$ takes a form which depends
on the complexity of the model for a given region of interest.
Between the discontinuities, each emission
region can be treated as having a single outer edge, a single inner
edge, or both an inner and outer edge.
For example, in MACS0744, the pre-shock region has a single inner edge,
the cold core has a single outer edge, and the shock-heated region has both and inner and outer
edge.
The total surface brightness is the sum of these three regions.

For one outer edge, we assume this edge is located at $r = 1$ in three dimensions and at $A = 1$ in projection.
Then, the bounds on the integral in equation~(\ref{equation:mainint2})
are $q_1=0$ and
\begin{equation}
q_2=c \left\{
\begin{array}{ll}
\left( 1-A^{2} \right)^{1/2} \, , & A < 1 \\
0 \, & A \ge 1 \, ,
\end{array}
\right.
\end{equation}
and $\phi$ takes the form
\begin{equation}
\phi=
\left\{
\begin{array}{ll}
1-I_{A^{2}}(p-\frac{1}{2},\frac{1}{2})\, , &  A < 1\\
0\, , & A \ge 1 \, .
\end{array}
\right.
\end{equation}
Here, $I_x (u,v)$ is the scaled incomplete beta function $I_x (u,v) \equiv B_x (u,v) / B(u,v)$,
$B_x (u,v) $ is the incomplete beta function, and $B(u,v) \equiv \Gamma(u) \Gamma(v) /  \Gamma(u+v)$ is the beta function.
Note that very efficient algorithms for calculating $B(u,v)$ and $I_x (u,v)$ exist and can be found as intrinsic functions on most computer systems.
Alternatively, they are given in {\it Numerical Recipes}
\citep{PTV+93}.

For a single inner edge located at $r=R$ in three dimension and at $A = R$ in projection,
the bounds are
\begin{equation}
q_1 = c \left\{
\begin{array}{ll}
\left( R^2-A^{2} \right)^{1/2} \, , & A < R \\
0 \, , & A \ge R \, ,
\end{array}
\right.
\end{equation}
and $q_2=\infty$
and we have
\begin{equation}
\phi=
\left\{
\begin{array}{ll}
I_{\frac{A^2}{R^2}}(p-\frac{1}{2},\frac{1}{2})\, , & A < R \\
1 \, , & A \ge R \, .
\end{array}
\right.
\end{equation}
It is useful to note that our expression for a single outer edge is
mathematically identical to the expression derived in \citet{vik01}
but uses the incomplete beta function (which is more convenient numerically) as opposed to the hyper-geometric function . 

Finally, for a region with two edges, we will take their locations to be $r = 1$ in three dimensions and $A = 1$ in projection for the inner edge, and $r = R$ or $A=R$ for the outer edge, where $R > 1$.
The bounds on the integral become
\begin{equation}
q_1 = c \left\{
\begin{array}{ll}
\left( 1-A^{2} \right)^{1/2} \, , & A < 1 \\
0 \, , & A \ge 1 \, ,
\end{array}
\right.
\end{equation}
and
\begin{equation}
q_2=c \left\{
\begin{array}{ll}
\left( R^2-A^{2} \right)^{1/2} \, , & A < R \\
0 \, , & A \ge R \, .
\end{array}
\right.
\end{equation}
The expression for $\phi$ becomes
\begin{equation}
\phi =
\left\{
\begin{array}{ll}
I_{A^{2}}(p-\frac{1}{2},\frac{1}{2}) - I_{\frac{A^{2}}{R^{2}}}(p-\frac{1}{2},\frac{1}{2}) \, ,
& A < 1 \\
1-I_{\frac{A^{2}}{R^{2}}}(p-\frac{1}{2},\frac{1}{2}) \, , & 1 \le A < R \\
0 \, , & A \ge R \, .
\end{array}
\right.
\end{equation}

\subsection{A.2: Density Profiles}

Once we have obtained the power law index $p$ and the normalization
$\varepsilon_{o}$ by fitting equation \ref{equation:model} to the
data, we can reconstruct the intrinsic emissivity distribution.
This is related to the density distribution $n_{e}(r)$ by
\begin{equation}
n_{e}(r)=\left[ \frac{\varepsilon(r)}{\Lambda (T_e,Z)} \right]^{1/2} \, ,
\end{equation}
where $\Lambda$ is the X-ray emissivity function which depends on electron temperature
$T_e$ and abundance $Z$.

If
XSPEC\footnote{http://heasarc.nasa.gov/xanadu/xspec/}
is used to determine the temperatures in the emission regions, the same models can easily be used to determine the value of $\Lambda$.
This has the great advantage that the models, temperature, abundances, and instrument responses used for the spectral analysis will be completely consistent with those used to determine $n_e (r)$.
We assume here that the model is a single-temperature MEKAL or APEC model.
For this purpose, only the shape of the spectrum matters, not its normalization, so the region fit in XSPEC need not be identical to the region fit in the surface brightness analysis, as long as the spectral shape is assumed to be the same.
If the surface brightness $I$ is analyzed in energy units, then the procedure is to determine the X-ray flux $F$ of the spectral region in the same band and with the same instrument as used to fit the surface brightness.
If the surface brightness was corrected for absorption, then the absorbing column should first be set to zero.
One also needs to record the normalization of the thermal model, which is defined as
\begin{equation}
K \equiv \frac{10^{-14}}{4 \pi ( 1 + z_r )^2 D_A^2} \int n_e n_p \, dV \, ,
\label{equation:xspecnorm}
\end{equation}
where $D_A$ is the angular diameter distance to the cluster, $n_p$ is the proton number density, and $V$ is the volume of the emitting region.
Then, the relevant X-ray emissivity function is
\begin{equation}
\Lambda = \frac{F( 1 + z_r )^2 }{10^{14} K  (n_e / n_p ) } \, .
\end{equation}
Here, $n_e / n_p \approx 1.21$ is the ratio of the electron to proton number densities, and is essentially a constant for typical cluster temperatures and abundances.

If the surface brightness is determined in count units (as is typically the case with X-ray observations), then the
procedure is to set the observed energy band and instrument in XSPEC to the one used for the surface brightness measurements, and then type ``show'' to determine the model countrate $CR$.
Then, the emissivity function is
\begin{equation}
\Lambda = \frac{CR \, ( 1 + z_r )}{10^{14} K (n_e / n_p ) } \, .
\end{equation}

\subsection{A.3: Pressure and SZE}

With a three dimensional density model obtained through the above
procedure and measurements of $T_e$ from X-ray spectroscopy, one can
produce a three dimensional pressure model which can be used to
predict the observed SZ flux.  From the ideal gas law, the electron pressure is simply
\begin{equation}
P_{e}(r)=k_{B}n_{e}(r) T_e (r) \, .
\end{equation}
By integrating this along the line of sight, one can obtain a two dimensional map of the
Compton $y_C$ parameter
\begin{equation}
y_C (x,y)=\int \frac{P_{e}(r)\sigma_{T}}{m_{e}c_{l}^{2}}\,dz.
\label{equation:y}
\end{equation}
Here, $k_{B}$ is Boltzmann's constant, $\sigma_{T}$ and $m_{e}$ are the Thomson
cross section and mass of the electron respectively, and $c_{l}$ is
the speed of light. 

Assuming that $T_e (r)$ is either a constant or is a power-law function of the radius within each region,
the electron pressure will vary as a power-law of the elliptical radius, and
the same analytic expression (equation~\ref{equation:model}) can be used to determine $y_C (x,y)$.
One simply makes the substitution
\begin{equation}
\frac{1}{4\pi(1+z_{r})^{\eta}} \varepsilon \rightarrow \frac{P_{e}(r)\sigma_{T}}{m_{e}c_{l}^{2}} \, .
\end{equation}
From a map of $y_C$, it is straightforward to produce a
model SZE image.



\section{B: Thermal Energy from SZE}
\label{section:etherm}

The surface brightness of a cluster due to the thermal SZE can be expressed, 
for dimensionless frequency  $x_{\nu} \equiv h\nu/\kB \Tcmb$ 
where $h$ is Planck's constant, $\nu$ is frequency, and 
\Tcmb\ is the primary CMB temperature, 
as the change $\Delta I_{\rm SZE}$ relative to the primary CMB surface brightness 
normalization $I_0$, as
\begin{eqnarray}
\label{eq:thermal_sz}
\frac{\Delta I_{\rm SZE}}{I_0} &=& \frac{\sigT}{m_e c_l^2} \int \!\!  g(x_{\nu},T_e) \, \kB n_e T_e \,dz \\
\label{eq:thermal_sz2}
&=& \frac{\sigT}{m_e c_l^2} \int \!\! g(x_{\nu},T_e) \, P_e \,dz \\
\label{eq:thermal_sz3}
&\equiv& g(x_{\nu},T_e) ~ y_C.
\end{eqnarray}
The primary CMB surface brightness normalization (in units of flux per solid angle) is
$I_0 = 2 (\kB \Tcmb)^3(h c_l)^{-2} = 2.7033 \times 10^8~\rm Jy/Sr$ 
[see e.g.\ \citet{chr}].
The factor $g(x_{\nu},T_e)$ encapsulates the SZE flux spectral dependence, which is a
function of electron temperature when relativistic corrections are taken into consideration.
In the classical physics limit, 
\begin{equation}
\label{eq:g_x}
g(x_{\nu}) =  \frac{x_{\nu}^4 e^{x_{\nu}}}{(e^{x_{\nu}}-1)^2} \left(x_{\nu} \frac{e^{x_{\nu}} + 1}{e^{x_{\nu}} - 1} - 4\right).
\end{equation}

We integrate the SZE surface brightness in Eq.\ \ref{eq:thermal_sz} to relate the
SZE flux from a region of the sky to the underlying electron pressure
in the measured ICM feature. 
The integrated SZE flux is computed (using Eqns. \ref{eq:thermal_sz2}, \ref{eq:thermal_sz3},
and \ref{equation:y})
\begin{equation}
\label{eq:Fsz}
F_{\rm SZE} 
= \int \!  I_{\rm SZE} \, d\Omega 
= g(x_{\nu}) {I_0} \int \! y_C  \, d\Omega 
= \frac{\sigT}{m_e c_l^2} \, g(x_{\nu}) {I_0} \int \! d\Omega \int \! P_e \,dz.
\end{equation}
Since $d\Omega = d\aleph/ D_A^2(z_r)$, where $d\aleph$ is the area integration element,
 $F_{\rm SZE}$
physically relates to the thermal energy $E_{\rm th}$ content of the gas
within a cylindrical region of a cluster (of volume $\Delta \aleph \Delta z$).
The electron pressure $P_e$ relates to the total pressure $P_{\rm tot}$ by 
the the electron weighting factor $\mu_e \approx 1.17$ (assuming standard abundances) 
as $P_{\rm tot} = (1 + 1/\mu_e) P_e$.
In terms of the flux (Eq.~\ref{eq:Fsz}), the thermal energy content is
\begin{equation}
\label{eq:Ethermal_Fsz}
E_{\rm th} =\frac{3}{2} \frac{(1 + 1/\mu_e) \, m_e c_l^2 \, F_{\rm SZE} \, D_A^2(z_r)}
{\sigT I_0 \, g(x_{\nu})}
\end{equation}
For MUSTANG data at 90~GHz, and an assumed $\kB T_e = 10~\rm keV$ and $\mu_e=1.17$, this is:
\begin{equation}
\label{eq:Ethermal_Fsz2}
E_{\rm th} = |F_{\rm SZE}| D_A^2(z_r) \rm\left[\frac{3.9\times10^{55} ergs}{mJy \, Mpc^2}\right]
\end{equation}
for $D_A(z_r)$ in Mpc.  
In this work, we use the \citet{itoh1998} relativistic corrections to Eq.~\ref{eq:g_x}.  

\bibliography{clusters_draft}

\end{document}